\documentclass[aps,pre,twocolumn,groupedaddress,amsmath,amssymb,floatfix]{revtex4}

\usepackage{graphicx}

\newcommand{\uu}{\mbox{\boldmath$u$}}
\newcommand{\vv}{\mbox{\boldmath$v$}}
\newcommand{\xx}{\mbox{\boldmath$x$}}
\newcommand{\D}{{\mathrm{d}}}

\begin{document}

\title{Stability and stabilisation of the \\ lattice Boltzmann method
 \\ Magic steps and salvation operations
}
\author{R. A. Brownlee}
\email[corresponding author: ]{r.brownlee@mcs.le.ac.uk}
\author{A. N. Gorban}
\author{J. Levesley}
\affiliation{Department of Mathematics, University of Leicester,
Leicester LE1 7RH, UK}

\begin{abstract}
We revisit the classical stability versus accuracy dilemma for the
lattice Boltzmann methods (LBM). Our goal is a stable method of
second-order accuracy for fluid dynamics based on the lattice
Bhatnager--Gross--Krook method (LBGK).

The LBGK scheme can be recognised as a discrete dynamical system
generated by free-flight and entropic involution. In this framework
the stability and accuracy analysis are more natural. We find the
necessary and sufficient conditions for second-order accurate fluid
dynamics modelling. In particular, it is proven that in order to
guarantee second-order accuracy the distribution should belong to a
distinguished surface -- the invariant film (up to second-order in
the time step). This surface is the trajectory of the
(quasi)equilibrium distribution surface under free-flight.

The main instability mechanisms are identified. The simplest recipes
for stabilisation add no artificial dissipation (up to second-order)
and provide second-order accuracy of the method. Two other
prescriptions add some artificial dissipation locally and prevent
the system from loss of positivity and local blow-up. Demonstration
of the proposed stable LBGK schemes are provided by the numerical
simulation of a $1$D shock tube and the unsteady $2$D-flow around a
square-cylinder up to Reynolds number $\mathcal{O}(10000)$.

\end{abstract}
\maketitle

\section{Introduction\label{sec1}}

A lattice Boltzmann method (LBM) is a discrete velocity method in
which a fluid is described by associating, with each velocity
$\vv_i$, a single-particle distribution function $f_i=f_i(\xx,t)$
which is evolved by advection and interaction on a fixed
computational lattice.

The method has been proposed as a discretization of Boltzmann's
kinetic equation (for an introduction and historic review see
\cite{succi01}). Furthermore, the collision operator can be
alluringly simplified, as is the case with the
Bhatnager--Gross--Krook (BGK) operator~\cite{bgk54}, whereby
collisions are described by a single-time relaxation to local
equilibria $f_i^*$:
\begin{equation}\label{lbgk}
\frac{\partial f_i}{\partial t}+\vv_i \cdot \nabla f_i  =
    \frac{1}{\tau}(f_i^*-f_i) .
\end{equation}
The physically reasonable choice for $f_i^*$ is as entropy
maximizers, although other choices of equilibria are often
preferred~\cite{succi01}. The local equilibria $f_i^*$ depend
nonlinearly on the hydrodynamic moments (density, momentum, etc.).
These moments are linear functions of $f_i$, hence~\eqref{lbgk} is a
nonlinear equation. For small $\tau$, the Chapman--Enskog
approximation~\cite{Chapman} reduces~\eqref{lbgk} to the
compressible Navier--Stokes equation~\cite{succi01} with kinematic
viscosity $\nu \sim \tau c_1^2$, where $c_1$ is the thermal velocity
for one degree of freedom.

The overrelaxation discretization of~\eqref{lbgk} (see,
e.g.,~\cite{Benzi,LB1,Higuera,succi01,karlin06,karlin07}) is known
as LBGK, and allows one to choose a time step $\delta_t \gg \tau$.
This decouples viscosity from the time step, thereby suggesting
that LBGK is capable of operating at arbitrarily high-Reynolds
number by making the relaxation time sufficiently small. However,
in this low-viscosity regime, LBGK suffers from numerical
instabilities which readily manifest themselves as local blow-ups
and spurious oscillations.

Another problem is the degree of accuracy. An approximation to the
continuous-in-time kinetics is not equivalent to an approximation of
the macroscopic transport equation. The fluid dynamics appears as a
singular limit of the Boltzmann or BGK equation for small $\tau$. An
approximation to the corresponding slow manifold in the distribution
space is constructed by the Chapman--Enskog expansion. This is an
asymptotic expansion, and higher (Burnett) terms could have
singularities. An alternative approach to asymptotic expansion (with
``diffusive scaling" instead of ``convective scaling" in the
Chapman--Enskog expansion) was developed in~\cite{Junk} in order to
obtain the incompressible Navier--Stokes equations directly from
kinetics.

It appears that the relaxation time of the overrelaxation scheme to
the slow hydrodynamic manifold may be quite large for small
viscosity: $t_{\rm relax} \sim c_1^2 \delta_t^2 /( 2 \nu)
\sim\delta_t^2 /( 2\tau)$ (see below, in Sec.~\ref{sec3}). Some
estimates of long relaxation time for LBGK at large Reynolds number
are found earlier in~\cite{Xiong}. So, instead of fast relaxation to
a slow manifold in continuous-in-time kinetics, we could meet a slow
relaxation to a fluid dynamics manifold in the chain of discrete LBM
steps.

Our approach is based on two ideas: the Ehrenfests'
coarse-graining~\cite{ehrenfest11,GKOeTPRE2001,gorban06} and the
method of differential approximation of difference
equations~\cite{Hirt,Shokin}. The background knowledge necessary to
discuss the LBM in this manner is presented in Sect.~\ref{sec2}. In
this section, we answer the question: how to provide second-order
accuracy of the LBM methods for fluid dynamics modelling? We prove
the necessary and sufficient conditions for this accuracy. It
requires a special connection between the distribution $f_i$ and the
hydrodynamic variables. There is only one degree of freedom for the
choice of $f_i$, if the hydrodynamic fields are given. Moreover, the
LBM with overrelaxation can provide approximation of the macroscopic
equation even when it does not approximate the continuous-in-time
microscopic kinetics.

This approach suggests several sources of numerical instabilities in
the LBM and allows several recipes for stabilisation. A geometric
background for this analysis provides a manifold that is a
trajectory $\mathfrak{q}$ of the quasiequilibrium manifold due to
free-flight. We call this manifold the \textit{invariant film}
(\textit{of nonequilibrium states}). It was introduced
in~\cite{ocherki} and studied further
in~\cite{gorban06,Plenka,GorKar}. Common to each stabilisation
recipe is the desire to stay uniformly close to the aforementioned
manifold (Sect.~\ref{sec3}).

In Sect.~\ref{sec5}, in addition to two LBM accuracy tests, a
numerical simulation of a $1$D shock tube and the unsteady $2$D-flow
around a square-cylinder using the present stabilised LBM are
presented. For the later problem, the simulation quantitatively
validates the experimentally obtained Strouhal--Reynolds
relationship up to $\mathrm{Re}=\mathcal{O}(10000)$. This extends
previous LBM studies of this problem where the relationship had only
been successfully validated up to
$\mathrm{Re}=\mathcal{O}(1000)$~\cite{ansumali04,baskar04}.

Sect.~\ref{sec6} contains some concluding remarks as well as
practical recommendations for LBM realisations.

We use operator notation that allows us to present general results
in compact form. The only definition we have to recall here is the
(G\^{a}teaux) differential: the differential of a map $J(f)$ at a
point $f_0$ is a linear operator $(D_f J)_{f_0}$ defined by a rule:
$(D_f J)_{f_0}g= \frac{\D}{\D \varepsilon}(J(f_0+ \varepsilon
g))_{\varepsilon=0}$.

\section{Background\label{sec2}}

\paragraph{Microscopic and macroscopic variables.}
Let us describe the main elements of the LBM construction. The first
element is a microscopic description,  a single-particle
distribution function $f(\xx,\vv)$, where $\xx$ is the space vector,
and $\vv$ is velocity. If velocity space is approximated by a finite
set $\{\vv_i\}$, then the distribution is approximated by a measure
with finite support, $f(\xx,\vv)\approx \sum_i f_i
\delta(\vv-\vv_i)$. In that case, the microscopic description is the
finite-dimensional vector-function $f_i(\xx)$.

The second main element is the macroscopic description. This is a
set of macroscopic vector fields that are usually some moments of
the distribution function. The main example gives the hydrodynamic
fields (density--momentum--energy density): $\{n, n \uu,
\mathcal{E}\}(\xx)=\int \{1, \vv, v^2/2\} f(\xx,\vv) \, \D  \vv$.
But this is not an obligatory choice. If we would like to solve by
LBM methods the Grad equations~\cite{Grad,HudongGrad} or some
extended thermodynamic equations~\cite{EIT}, we should extend the
list of moments (but, at the same time, we should be ready to
introduce more discrete velocities for a proper description of
these extended moment systems).

In general, we use the notation $f$ for the microscopic state, and
$M$ for the macroscopic state. The vector $M$ is a linear function
of $f$: $M=m(f)$.

\paragraph{Equilibrium.}
For any allowable value of $M$ an ``equilibrium'' distribution
should be given: a microscopic state $f^*_M$. It should satisfy the
obvious, but important identity of self-consistency:
\begin{equation}\label{SelfConsId}
m(f^*_M)=M,
\end{equation}
or in differential form
\begin{equation}\label{SelfConsIdDif}
m(D_Mf^*_M)\equiv1,\quad\text{i.e., $m((D_Mf^*_M)a)\equiv a$}.
\end{equation}
The state $f^*_M$ is not a proper thermodynamic equilibrium, but a
conditional one under the constraint $m(f)=M$. Therefore we call it
a \textit{quasiequilibrium} (other names, such as local equilibrium,
conditional equilibrium, generalised canonical state or
pseudoequilibrium are also in use).

For the quasiequilibrium $f^*_M$, an equilibration operation is the
projection $\Pi^*$ of the distribution $f$ into the corresponding
quasiequilibrium state: $\Pi^*(f)=f^*_{m(f)}$.

In the fully physical situation with continuous velocity space, the
quasiequilibrium $f^*_M$ is defined as a conditional entropy
maximizer by a solution of the optimisation problem:
\begin{equation}\label{smax}
S(f)\to \max,\ m(f)=M,
\end{equation}
where $S(f)$ is an entropy functional.

The choice of entropy is ambiguous; generally, we can start from a
concave functional of the form
\begin{equation}\label{entropy}
    S(f)=\int s(f(\xx,\vv,t)) f(\xx,\vv,t) \, \D \xx \D \vv
\end{equation}
with a concave function of one variable $s(f)$. The choice by
default is $s(f)=-\ln f$, which gives the classical
Boltzmann--Gibbs--Shannon (BGS) entropy.

For discrete velocity space, there exist some extra moment
conditions on the equilibrium construction: in addition to
\eqref{SelfConsId} some higher moments of a discrete equilibrium
should be the same as for the continuous one. This is necessary to
provide the proper macroscopic equations for $M$. Existence of
entropy for the entropic equilibrium definition~\eqref{smax} whilst
fulfilling higher moment conditions could be in contradiction, and a
special choice of velocity set may be necessary (for a very recent
example of such research for multispeed lattices see
\cite{Shyam2006}). Another choice is to refuse to deal with the
entropic definition of equilibrium~\eqref{smax} and assume that
there will be no perpetuum mobile of the second kind. This extends
the possibility for approximation, but creates some risk of
nonphysical behavior of the model. For a detailed discussion of the
$H$-theorem for LBM we refer the readers to~\cite{LB3}.

Some of the following results depend on the entropic definition of
equilibrium, but some do not. We always point out if results are
``entropy--free".

\paragraph{Free flight.}

In the LBM construction the other main elements are: the free-flight
transformation and the collision. There are many models of
collisions, but the free-flight equation is always the same
\begin{equation}\label{freeflight}
\frac{\partial f}{\partial t}+\vv \cdot \nabla_x f=0,
\end{equation}
with exact solution $f(\xx,\vv,t)=f(\xx-\vv t, \vv, 0)$, or for
discrete velocities,
\begin{equation}\label{ffdisc}
\frac{\partial f_i}{\partial t}+\vv_i \cdot \nabla_x f_i  =0,
\end{equation}
$f_i (\xx,t)=f_i(\xx-\vv_i t,0)$. Free-flight conserves any entropy
of the form~\eqref{entropy}. In general, we can start from any
dynamics. For application of the entropic formalism, this dynamics
should conserve entropy. Let this kinetic equation be
\begin{equation}\label{conskin}
\frac{\D f}{\D t} = J_c(f).
\end{equation}
For our considerations, the free-flight equation will be the main
example of the conservative kinetics~\eqref{conskin}.

The phase flow $\Theta_{t}$ for kinetic equation~\eqref{conskin} is
a shift in time that transforms $f(t_0)$ into $f(t_0+t)$. For
free-flight, $\Theta_t: f(\xx,\vv) \to f(\xx-\vv t,\vv)$.

\paragraph*{Remark.}
We work with dynamical systems defined by partial
differential equations. Strictly speaking, this means that the
proper boundary conditions are fixed. In order to separate the
discussion of equation from a boundary condition problem, let us
imagine a system with periodic boundary conditions (e.g., on a
torus), or a system with equilibrium boundary conditions at
infinity.

\paragraph{Ehrenfests' solver of second-order accuracy for the Navier--Stokes equations.}

Here we present a generalisation of a well known result. Let us
study the following process (an example of the Ehrenfests' chain
\cite{ehrenfest11,gorban06,GKOeTPRE2001}, a similar result gives the
optimal prediction approach~\cite{Raz}): free-flight for time $\tau$
-- equilibration -- free-flight for time $\tau$ -- equilibration --
$\dotsb$. During this process, the hydrodynamic fields approximate
the solution of the (compressible) Navier--Stokes equation with
viscosity $\nu \sim \frac{\tau}{2} c_1^2$, where $c_1$ is the
thermal velocity for one degree of freedom. The error of one step of
this approximation has the order $\mathcal{O}(\tau^3)$. An exact
expression for the transport equation that is approximated by this
process in the general situation (for arbitrary initial kinetics,
velocity set and for any set of moments) is:
\begin{equation}\label{MACRO1}
\frac{\D M}{\D t} = m(J_c(f^*_M)) +\frac{\tau}{2}m((D_f
J_c(f))_{f^*_M}\Delta_{f^*_M}),
\end{equation}
where $\Delta_{f^*_M}$ is the \textit{defect of invariance} of the
quasiequilibrium manifold:
\begin{equation}\label{defect}
\Delta_{f^*_M}=J_c(f^*_M)- D_M (f^*_M)m(J_c(f^*_M)),
\end{equation}
and is the difference between the vector-field $J_c$ and its
projection on to the quasiequilibrium manifold. This result is
entropy-free.

The first term in the right hand side of~\eqref{MACRO1} --
\textit{the quasiequilibrium approximation} -- consists of moments
of ${\D f}/{\D t}$ computed at the quasiequilibrium point. For
free-flight, hydrodynamic fields and Maxwell equilibria this term
gives the Euler equations. The second term includes the action of
the differential $D_f J_c(f)_{f^*_M}$ on the defect of invariance
$\Delta_{f^*_M}$ (for free-flight~\eqref{freeflight}, this
differential is just $-\vv \cdot \nabla_x$, for the discrete version
\eqref{ffdisc} this is the vector-column $-\vv_i \cdot \nabla_x$).
These terms always appear in the Chapman--Enskog expansion. For
free-flight, hydrodynamic fields and Maxwell equilibria they give
the Navier--Stokes equations for a monatomic gas with Prandtl number
$\mathrm{Pr}=1$:
\begin{equation}
\begin{split}\label{NSEq}
\frac{\partial n}{\partial t}&=-\sum_i \frac{\partial (nu_i)}{\partial x_i},\\
\frac{\partial (nu_k)}{\partial t}&=-\sum_i\frac{\partial
(nu_ku_i)}{\partial x_i}-\frac{1}{\rm m}\frac{\partial P}{\partial x_k} \\
&+ \underline{\frac{\tau}{2}\frac{1}{\rm m} \sum_i
\frac{\partial}{\partial x_i}\left[P\left(\frac{\partial
u_k}{\partial x_i}+\frac{\partial u_i}{\partial x_k}-\frac{2}{3}
\delta_{ki}{\rm div} u \right)\right]},  \\
\frac{\partial \mathcal{E}}{\partial t} &=-\sum_i\frac{\partial
(\mathcal{E} u_i)}{\partial x_i} - \frac{1}{\rm m}
\sum_i\frac{\partial (Pu_i)}{\partial x_i}  \\&\qquad +
\underline{\frac{\tau}{2}\frac{5k_{\rm B}}{2{\rm
m}^2}\sum_i\frac{\partial}{\partial x_i}\left(P\frac{\partial T}{
\partial x_i}\right)},
\end{split}
\end{equation}
where $\rm m$ is particle mass, $k_B$ is Boltzmann's constant, $P=
nk_{\rm B}T$ is ideal gas pressure, $T$ is kinetic temperature, and
the underlined terms are the results of the coarse-graining
additional to the quasiequilibrium approximation.

All computations are straightforward exercises (differential
calculus and Gaussian integrals for computation of the moments, $m$,
in the continuous case). More details of these computations are
presented in~\cite{GorKar}.

The dynamic viscosity in~\eqref{NSEq} is $\mu =\frac{\tau}{2}n
k_{\rm B }T$. It is useful to compare this formula to the
mean-free-path theory that gives $\mu = \tau_{\rm col} n k_{\rm B
}T$, where  $\tau_{\rm col}$ is the collision time (the time for the
mean-free-path). According to these formulas, we get the following
interpretation of the coarse-graining time $\tau$ for this example:
$\tau=2 \tau_{\rm col}$.

For any particular choice of discrete velocity set $\{\vv_i \}$ and
of equilibrium $f^*_M$ the calculation could give different
equations, but the general formula~\eqref{MACRO1} remains the same.
The connection between discretization and viscosity was also studied
in~\cite{SChen}. Let us prove the general formula~\eqref{MACRO1}.

We are looking for a macroscopic system that is approximated by the
Ehrenfests' chain. Let us look for macroscopic equations of the form
\begin{equation}
\frac{\D M}{\D t}=\Psi(M)
\end{equation}
with the phase flow $\Phi_t$: $M(t)=\Phi_t M(0)$. The transformation
$\Phi_{\tau}$ should coincide with the transformation $M\mapsto
m(\Theta_{\tau}(f^*_M))$ up to second-order in $\tau$. The matching
condition is
\begin{equation}\label{match2}
m(\Theta_{\tau}(f^*_M))=\Phi_{\tau}(M) \quad \text{for every $M$ and
given $\tau$}.
\end{equation}
This condition is the equation for the macroscopic vector field
$\Psi(M)$. The solution of this equation is a function of $\tau$:
$\Psi=\Psi(M,\tau)$. For a sufficiently smooth microscopic vector
field $J_c(f)$ and entropy $S(f)$ it is easy to find the Taylor
expansion of $\Psi(M,\tau)$ in powers of $\tau$. Let us find the
first two terms: $\Psi(M,\tau)=\Psi_0(M)+\tau \Psi_1(M)+o(\tau)$. Up
to second-order in $\tau$ the matching condition~\eqref{match2} is
\begin{multline}\label{match3}
m(J_c(f^*_M))\tau+m((D_f J_c(f))_{f^*_M} (J_c(f^*_M)))\frac{\tau^2}{2} \\
=\Psi_0(M)\tau+ \Psi_1(M)\tau^2  + (D_M
\Psi_0(M))(\Psi_0(M))\frac{\tau^2}{2}.
\end{multline}

From this condition immediately follows:
\begin{equation}
\begin{split}\label{CoaGr2} \Psi_0(M) &= m(J_c(f^*_M)),\\
\Psi_1(M)&=\frac{1}{2}m((D_f J_c(f))_{f^*_M} \Delta_{f^*_M}),
\end{split}
\end{equation}
where $\Delta_{f^*_M}$ is the defect of invariance~\eqref{defect}.
Thus we find that the macroscopic equation in the first
approximation is~\eqref{MACRO1}.

\paragraph{The Chapman--Enskog expansion for the generalised BGK
equation.}

Here we present the Chapman--Enskog method for a class of
generalised model equations. This class includes the well-known BGK
kinetic equation, as well as many other model
equations~\cite{GKMod}.

As a starting point we take a formal kinetic equation with a small
parameter $\tau$
\begin{equation}\label{FormSingPert}
\frac{\D f}{\D t} = J(f) := J_c(f)+\frac{1}{\tau}(\Pi^*(f)- f).
\end{equation}
The term $\Pi^*(f)- f$ is nonlinear because of the nonlinear
dependency of $\Pi^*(f)=f^*_{m(f)}$ on $m(f)$.

We would like to find a reduced description valid for the
macroscopic variables $M$. This means, at least, that we are looking
for an invariant manifold parameterised by $M$, $f=f_M$, that
satisfies the {\it invariance equation}:
\begin{equation}\label{InvEq}
(D_M f_M)( m(J(f_M)))=J(f_M).
\end{equation}

The invariance equation means that the time derivative of $f$
calculated through the time derivative of $M$ ($\dot{M} =m(J(f_M))$)
by the chain rule coincides with the true time derivative $J(f)$.
This is the central equation for model reduction theory and
applications. The first general results about existence and
regularity of solutions to~\eqref{InvEq} were obtained by Lyapunov
\cite{Lya} (see, e.g., the review in~\cite{GorKar}). For the kinetic
equation~\eqref{FormSingPert} the invariance equation has the form
\begin{equation}\label{InvEqSing}
(D_M f_M)( m(J_c(f_M)))=J_c(f_M)+\frac{1}{\tau}(f^*_{M}- f_M),
\end{equation}
because of the self-consistency identity~\eqref{SelfConsId},
\eqref{SelfConsIdDif}.

Due to the presence of the small parameter $\tau$ in $J(f)$, the
zeroth approximation to $f_M$ is the quasiequilibrium approximation:
$f^{(0)}_M=f^*_M$. Let us look for $f_M$ in the form of a power
series: $f_M =f^{(0)}_M+{\tau}f^{(1)}_M+\dotsb$, with
$m(f^{(k)}_M)=0$ for $k\geq 1$. From~\eqref{InvEqSing} we
immediately find:
\begin{equation}\label{firstChEnBGKmic}
f^{(1)}_M = J_c(f^{(0)}_M) - (D_M f^{(0)}_M)( m(J_c(f^{(0)}_M)))=
\Delta_{f^*_M}.
\end{equation}
It is very natural that the first term of the Chapman--Enskog
expansion for the model equation~\eqref{FormSingPert} is just the
defect of invariance for the quasiequilibrium~\eqref{defect}.

The corresponding first-order in $\tau$ approximation for the
macroscopic equations is:
\begin{equation}\label{firstChEnBGKmac}
\frac{\D M}{\D t}=m(J_c(f^*_M))+ \tau m((D_f
J_c(f))_{f^*_M}\Delta_{f^*_M}).
\end{equation}
We should recall that $m(\Delta_{f^*_M})=0$. The last term in
\eqref{InvEqSing} vanishes in the macroscopic projection for all
orders. The only difference between~\eqref{firstChEnBGKmac} and
\eqref{MACRO1} is the coefficient 1/2 before $\tau$ in
\eqref{MACRO1}.

\paragraph{Decoupling of time step and viscosity:
how to provide second-order accuracy?}

In the Ehrenfests' chain ``free-flight -- equilibration -- $\dotsb$"
the starting point of each link is a quasiequilibrium state: the
chain starts from $f^*_{M(0)}$, then, after free-flight,
equilibrates into $f^*_{M(\tau)}$, etc. The viscosity coefficient in
\eqref{MACRO1} is proportional to $\tau$. Let us choose another
starting point $f^s_M$ in order to decouple time step and viscosity
and preserve the second-order accuracy of approximation. We would
like to get equation~\eqref{MACRO1} with a chain time step
$\delta_t= h$. Analogously to~\eqref{match3} and~\eqref{CoaGr2}, we
obtain the macroscopic equation
\begin{equation}\label{MACRO3}
\frac{\D M}{\D t}=m(J_c(f^*_M))\\ + m((D_f J_c(f))_{f^*_M}((f^s_M -
f^*_M) + ({h}/{2}) \Delta_{f^*_M})),
\end{equation}
under the condition that $f^s_M - f^*_M = \mathcal{O}(h)$. The
initial point
\begin{equation}\label{incond}
f^s_M = f^*_M -\frac{1}{2}(h-\tau)\Delta_{f^*_M} + {o}(h)
\end{equation}
provides the required viscosity. This is a sufficient condition for
the second-order accuracy of the approximation. Of course, the
self-consistency identity $m(f^s_M)=M$ should be valid exactly, as
\eqref{SelfConsId} is. This starting distribution is a linear
combination of the quasiequilibrium state and the first
Chapman--Enskog approximation.

The necessary and sufficient condition for second-order accuracy of
the approximation is:
\begin{equation}\label{incond1}
m\left((D_f J_c(f))_{f^*_M}( f^s_M - f^*_M +
\frac{1}{2}(h-\tau)\Delta_{f^*_M})\right) = {o}(h)
\end{equation}
(with the self-consistency identity $m(f^s_M)=M$). This means, that
the difference between left and right hand sides of~\eqref{incond}
should have zero moments and give zero inputs in observable
macroscopic fluxes.

Hence, the condition of second-order accuracy significantly
restricts the possible initial point for free-flight. This result is
also entropy-free.

Any construction of collisions should keep the system's starting
free-flight steps near the points $f^s_M$ given by~\eqref{incond}
and~\eqref{incond1}. The conditions~\eqref{incond} and
\eqref{incond1} for second-order accuracy of the transport equation
approximation do not depend on a specific collision model, they are
valid for most modifications of the LBM and LBGK that use
free-flight as a main step.

Various multistep approximations give more freedom of choice for the
initial state. For the construction of such approximations below,
the following {\it mean viscosity lemma} is important: if the
transformations $\Omega _i^h: M \rightarrow M'$,  $i=1,\ldots,k$,
approximate the phase flow for~\eqref{MACRO1} for time $h$ (shift in
time $h$) and $\tau=\tau_i$ with second-order accuracy in $h$, then
the superposition $\Omega _1^h \Omega _2^h\dotsb \Omega _k^h$
approximates the phase flow for~\eqref{MACRO1} for time $kh$ (shift
in time $kh$) for the average viscosity $\tau=\frac{1}{k}(\tau_1 +
\dotsb +\tau_k)$ with the same order of accuracy. The proof is by
straightforward multiple applications of Taylor's formula.

\paragraph{Entropic formula for $D_M (f^*_M)$.}
Among the many benefits of thermodynamics for stability analysis
there are some technical issues too. The differential of equilibrium
$D_M (f^*_M)$ appears in many expressions, for example
\eqref{SelfConsIdDif},~\eqref{defect},~\eqref{match3},
\eqref{CoaGr2},~\eqref{InvEq} and~\eqref{InvEqSing}. If the
quasiequilibrium is defined by the solution of the optimisation
problem~\eqref{smax}, then
\begin{equation}\label{QEP0}
D_Mf^*_M=\left(D_f^2S \right)_{f^*_{M}}^{-1}m^T
\left(m\left(D_f^2S\right)_{f^*_{M}}^{-1}m^T\right)^{-1}.
\end{equation}
This operator is constructed from the vector $m$, the transposed
vector $m^T$ and the second differential of entropy. The inverse
Hessian $(\partial^2 S  / \partial f_i \partial f_j)^{-1}$ is
especially simple for the BGS entropy, it is just $f_i \delta_{ij}$.
The formula~\eqref{QEP0} was first obtained in~\cite{Robertson} (for
an important particular case; for further references see
\cite{GorKar}).

\paragraph{Invariant film.}
All the points $\Theta_{t}(f^*_M)$ belong to a manifold that is a
trajectory $\mathfrak{q}$ of the quasiequilibrium manifold due to
the conservative dynamics~\eqref{conskin} (in hydrodynamic
applications this is the free-flight dynamics~\eqref{freeflight}. We
call this manifold the \textit{invariant film} (\textit{of
nonequilibrium states}). It was introduced in~\cite{ocherki} and
studied further in~\cite{gorban06,Plenka,GorKar}. The defect of
invariance $\Delta_{f^*_M}$~\eqref{defect} is tangent to
$\mathfrak{q}$ at the point $f^*_M$, and belongs to the intersection
of this tangent space with $\ker m$. This intersection is
one-dimensional. This means that the direction of $\Delta_{f^*_M}$
is selected from the tangent space to $\mathfrak{q}$ by the
condition: derivative of $M$ in this direction is zero.

A point $f$ on the invariant film $\mathfrak{q}$ is naturally
parameterised by $(M,t)$: $f=q_{M,t}$, where $M=m(f)$ is the value
of the macroscopic variables, and $t = t (f)$ is the time shift from
a quasiequilibrium state: $\Theta_{-t }(f)$ is a quasiequilibrium
state for some (other) value of $M$. By definition, the action of
$\Theta_{t }$ on the second coordinate of $q_{M,t}$ is simple:
$\Theta_{t }(q_{M,\tau})=q_{M',t+\tau}$. To the first-order in $t$,
\begin{equation}\label{filmFirOr}
    q_{M,t}=f^*_M+ t \Delta_{f^*_M},
\end{equation}
and $q_{M,0}\equiv f^*_M$.  The quasiequilibrium manifold divides
$\mathfrak{q}$ into two parts, $\mathfrak{q} = \mathfrak{q}_- \cup
\mathfrak{q}_0 \cup \mathfrak{q}_+$, where $\mathfrak{q}_- =
\{q_{M,t}|\,t < 0 \}$, $\mathfrak{q}_+ = \{q_{M,t}|\,t > 0 \}$, and
$\mathfrak{q}_0$ is the quasiequilibrium manifold: $\mathfrak{q}_0 =
\{q_{M,0} \}= \{f^*_M\}$.

There is an important {\it temporal involution} of the film:
\begin{equation}\label{TempInv}
I_T(q_{M,t})=q_{M,-t}.
\end{equation}

Due to~\eqref{incond}, for $q_{M,t}$ and a given time step $h$ the
transformation $M \mapsto m(\Theta_{h}(q_{M,t}))$ approximates the
solution of~\eqref{MACRO1} with $\tau=2t+h$ for the initial
conditions $M$ and time step $h$ with second-order accuracy in $h$.
Hence, due to the mean viscosity lemma, the two-step transformation
\begin{equation}\label{TwoStep}
M\mapsto m (I_T (\Theta_{h} (I_T (\Theta_{h} (q_{M,t})))))
\end{equation}
approximates the solution of~\eqref{MACRO1} with $\tau=0$ (the Euler
equations) for the initial conditions $M$ and time step $h$ with
second-order accuracy in $h$. This is true for any $t$, hence, for
any starting point on the invariant film with the given value of
$M$.

To approximate the solution of~\eqref{MACRO1} with nonzero $\tau$,
we need an incomplete involution:
\begin{equation}\label{TempInvInc}
I_T^{\beta}(q_{M,t})=q_{M,-(2\beta -1)t}.
\end{equation}
For $\beta=1$, we have $I_T^{1}=I_T$ and for $\beta=1/2$,
$I_T^{1/2}$ is just the projection onto the quasiequilibrium
manifold: $I_T^{1/2}(q_{M,t})=\Pi^*(q_{M,t})= q_{M,0}$. After some
initial steps, the following sequence gives a second-order in time
step $h$ approximation of~\eqref{MACRO1} with
$\tau=(1-\beta)h/\beta$, $1/2\leq\beta\leq1$:
\begin{equation}\label{coup}
M_n = m((I_T^{\beta} \Theta_{h})^n q_{M,t}).
\end{equation}
To prove this statement we consider a transformation of the second
coordinate in $q_{M,\vartheta_n}$ by $I_T^{\beta} \Theta_{h}$:
\begin{equation}\label{jump}
\vartheta_{n+1} = -(2\beta - 1)(\vartheta_{n} + h).
\end{equation}
This transformation has a fixed point $\vartheta^* = - h (2\beta -
1)/(2\beta)$ and $\vartheta_n=\vartheta^* + (-1)^n (2\beta - 1)^n
\delta$ for some $\delta$. If $1-\beta$ is small then relaxation
may be very slow: $\vartheta_n \approx \vartheta^* + (-1)^n \delta
\exp(-2n(1-\beta))$, and relaxation requires $\sim 1/(2(1-\beta))$
steps. If $\vartheta_n = \vartheta^* + o(h)$ then the sequence
$M_n$~\eqref{coup} approximates~\eqref{MACRO1} with $\tau= h -
2|\vartheta^*| = (1-\beta)h / \beta$ and second-order accuracy in
the time step $h$. The fixed points $q_{M,\vartheta^*}$ coincide
with the restart points $f^*_M + \vartheta^* \Delta_{f^*_M}$
(\ref{incond}) in the first order in $\vartheta^*=-(h-\tau)/2$,
and the middle points $\vartheta^*+h/2$ of the free-flight jumps
$q_{M,\vartheta^*} \mapsto q_{M',\vartheta^*+h}$ approximate the
first-order Chapman--Enskog manifold $f^*_M + \frac{\tau}{2}
\Delta_{f^*_M}$.

For the entropic description of quasiequilibrium, we can connect
time with entropy and introduce entropic coordinates. For each $M$
and positive $s$ from some interval $0<s<\varsigma$ there exist two
numbers $t_{\pm}(M,s)$ ($t_+(M,s) >0$, $t_-(M,s) <0$) such that
\begin{equation}\label{Entropy-time}
    S(q_{M,t_{\pm}(M,s)})=S(f^*_M)-s.
\end{equation}
The numbers $t_{\pm}$ coincide to the first-order: $t_+=-t_- +
o(t_-)$.

We define the \textit{entropic involution} $I_S$ as a transformation
of $\mathfrak{q}$:
\begin{equation}\label{EntrInv}
I_S(q_{M,t_{\pm}})=q_{M,t_{\mp}}.
\end{equation}
The introduction of incomplete entropic involution $I_S^\beta$ is
also obvious (see~\cite{gorban06}).

Entropic involution $I_S$ coincides with the temporal involution
$I_T$, up to second-order in the deviation from quasiequilibrium
state $f-\Pi^*(f)$. Hence, in the vicinity of quasiequilibrium there
is no significant difference between these operations, and all
statements about the temporal involution are valid for the entropic
involution with the same level of accuracy.

For the transfer from free-flight with temporal or entropic
involution to the standard LBGK models we must transfer from
dynamics and involution on $\mathfrak{q}$ to the whole space of
states. Instead of $I_T^{\beta}$ or $I_S^{\beta}$  the
transformation
\begin{equation}\label{linInv}
    I_0^{\beta}:f \mapsto \Pi^*(f)+ (2\beta-1)(\Pi^*(f)-f)
\end{equation}
is used. For $\beta=1$, $I_0^{1}$ is a mirror reflection in the
quasiequilibrium state $\Pi^*(f)$, and for $\beta=1/2$, $I_0^{1/2}$
is the projection onto the quasiequilibrium manifold. If, for a
given $f_0=q_{M,t}$, the sequence~\eqref{coup} gives a second-order
in time step $h$ approximation of~\eqref{MACRO1}, then the sequence
\begin{equation}\label{coupLin}
    M_n = m((I_0^{\beta} \Theta_{h})^n f_0)
\end{equation}
also gives a second-order approximation to the same equation with
$\tau= (1-\beta)h / \beta$ . This chain is the standard LBGK
model.

Entropic LBGK (ELBGK)
methods~\cite{boghosian01,gorban06,KGSBPRL,LB3} differ only in the
definition of~\eqref{linInv}: for $\beta=1$ it should conserve
entropy, and in general has the following form:
\begin{equation}\label{elbm}
I_E^{\beta}(f)= (1-\beta)f+\beta \tilde{f},
\end{equation}
with $\tilde{f}= (1-\alpha) f+\alpha \Pi^*(f)$. The number
$\alpha=\alpha(f)$ is chosen so that a constant entropy condition is
satisfied: $S(f)=S(\tilde{f})$. For LBGK~\eqref{linInv}, $\alpha=2$.

Of course, computation of $I_0^\beta$ is much easier than that of
$I_T^\beta$, $I_S^\beta$ or $I_E^\beta$: it is not necessary to
follow exactly the manifold $\mathfrak{q}$ and to solve the
nonlinear constant entropy condition equation. For an appropriate
initial condition from $\mathfrak{q}$ (not sufficiently close to
$\mathfrak{q}_0$), two steps of LBGK with $I_0^{\beta}$ give the
same second-order accuracy as~\eqref{coup}. But a long chain of such
steps can lead far from the quasiequilibrium manifold and even from
$\mathfrak{q}$. Here, we see stability problems arising. For $\beta$
close to 1, the one-step transformation $I_0^{\beta} \Theta_{h}$ in
the chain~\eqref{coupLin} almost conserves distances between
microscopic distributions, hence, we cannot expect fast exponential
decay of any mode, and this system is near the boundary of Lyapunov
stability.

\paragraph{Does LBGK with overrelaxation collisions approximate the BGK
equation?} The BGK equation as well as its discrete velocity version
\eqref{lbgk} has a direction of fast contraction $\Pi^*(f)-f$. The
discrete chain~\eqref{coupLin} with $\beta$ close to 1 has nothing
similar. Hence, the approximation of a genuine BGK solution by an
LBGK chain may be possible only if both the BGK and the LBGK chain
trajectories belong to a slow manifold with high accuracy. This
implies significant restrictions on initial data and on the dynamics
of the approximated solution, as well as fast relaxation of the LBGK
chain to the slow manifold.

The usual Taylor series based arguments from~\cite{HeLuo} are valid
for $h \ll \tau$. If we assume $h \gg \tau$, ($\delta_t \gg \lambda$
in the notation of~\cite{HeLuo}) then Eqn.~(10) of~\cite{HeLuo}
transforms (in our notation) into $f(\xx+\vv h, \vv, t+h)
=f^*_M(\xx+\vv h, \vv, t+h) + \mathcal{O}(\tau)$ with $M=m(f(\xx+\vv
h, \vv, t+h))$. That is, $f(\xx, \vv, t+h) =f^*_M(\xx, \vv, t+h) +
\mathcal{O}(\tau)$. According to this formula, $f$ should almost be
at quasiequilibrium after a time step $h \gg \tau$, with some
correction terms of order $\tau$. This first-order in $\tau$
correction is, of course, the first term of the Chapman--Enskog
expansion~\eqref{firstChEnBGKmic}: $\tau f^{(1)}_M = \tau
\Delta_{f^*_M}$ (with possible error of order $\mathcal{O}(\tau
h))$. This is a very natural result for an approximation of the BGK
solution, especially in light of the Chapman--Enskog expansion
\cite{Chapman,HeLuo}, but it is not the LBM scheme with
overrelaxation.

The standard element in the proof of second-order accuracy of the
BGK equation approximation by an LBGK chain uses the estimation of
an integral: for time step $h$ we obtain from~\eqref{lbgk} the exact
identity
\begin{equation}\label{steping}
f_i (\xx+\vv_i h,t+h)=\frac{1}{\tau}\int_t^{t+h}(
f^*_{i,m(f)}(\xx)-f_i(\xx,t')) \D t',
\end{equation}
where $f^*_{i,m(f)}(\xx)$ is the quasiequilibrium state that
corresponds to the hydrodynamic fields $m(f(\xx,t'))$. Then one
could apply the trapezoid rule for integration to the right-hand
side of~\eqref{steping}. The error of the trapezoid rule has the
order $\mathcal{O}(h^3)$:
\begin{equation*}
\int_t^{t+h} Q(t') \, \D t'= \frac{h}{2}(Q(t)+Q(t+h))-
\frac{h^3}{12} \ddot{Q}(t'),
\end{equation*}
where $t' \in [t,t+h]$ is a priori unknown point. But for the
singularly perturbed system~\eqref{lbgk}, the second derivative of
the term $f^*_{i,m(f)}(\xx)-f_i(\xx,t')$ on the right hand side
of~\eqref{steping} could be of order $1/\tau^2$, and the whole error
estimate is $\mathcal{O}(h^3/\tau^3)$. This is not small for
$h>\tau$. For backward or forward in time estimates of the
integral~\eqref{steping}, errors have the order
$\mathcal{O}(h^2/\tau^2)$. Hence, for overrelaxation with $h\gg\tau$
this reasoning is not applicable. Many simple examples of
quantitative and qualitative errors of this approximation for a
singularly perturbed system could be obtained by analysis of a
simple system of two equations: $\dot{x}=\frac{1}{\tau}(\phi(y)-x)$,
$\dot{y}=\psi(x,y)$ for various $\phi$ and $\psi$. There are
examples of slow relaxation (instead of fast), of blow-up instead of
relaxation or of spurious oscillations, etc.

Hence, one cannot state that LBGK with overrelaxation collisions
approximates solutions of the BGK equation. Nevertheless, it can do
another job: it can approximate solutions of the macroscopic
transport equation. As demonstrated within this section, the LBGK
chain~\eqref{coupLin}, after some initial relaxation period,
provides a second-order approximation to the transport equation, if
it goes close to the invariant film up to the order
$\mathcal{O}(h^2)$ (this initial relaxation period may have the
order $\mathcal{O}(h^2/\tau)$). In other words, it gives the
required second-order approximation for the macroscopic transport
equation under some stability conditions.

\section{Stability and stabilisation\label{sec3}}

\subsection{Instabilities}

\paragraph{Positivity loss.}

First of all, if $f$ is far from the quasiequilibrium, the state
$I_0^{\beta}(f)$~\eqref{linInv} may be nonphysical. The positivity
conditions (positivity of probabilities or populations) may be
violated. For multi- and infinite-dimensional problems it is
necessary to specify what one means by \textit{far}. In the previous
section, $f$ is the whole state which includes the states of all
sites of the lattice. All the involution operators with classical
entropies are defined for lattice sites independently. Violation of
positivity at one site makes the whole state nonphysical. Hence, we
should use here the $\ell_{\infty}$-norm: close states are close
uniformly, at all sites.

\paragraph{Large deviations.}
The second problem is nonlinearity: for accuracy estimates we always
use the assumption that $f$ is sufficiently close to
quasiequilibrium. Far from the quasiequilibrium manifold these
estimates do not work because of nonlinearity (first of all, the
quasiequilibrium distribution, $f^*_M$, depends nonlinearly on $M$
and hence the projection operator, $\Pi^*$, is nonlinear). Again we
need to keep the states not far from the quasiequilibrium manifold.

\begin{figure}
\begin{centering}
\includegraphics[width=70mm, height=40mm]{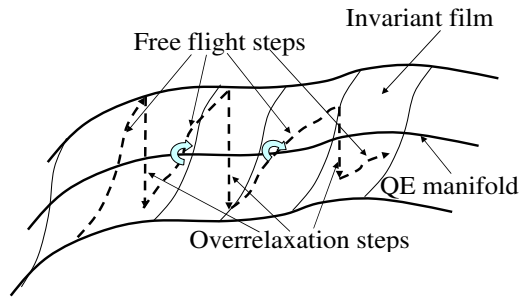}
\caption{\label{DirStab}Directional instability: after several
iterations the trajectory is not tangent to the invariant film with
the required accuracy.}
\end{centering}
\end{figure}

\paragraph{Directional instability.}
The third problem is a directional instability  that can affect
accuracy: the vector $f-\Pi^*(f)$ can deviate far from the tangent
to $\mathfrak{q}$ (Fig.~\ref{DirStab}). Hence, we should not only
keep $f$ close to the quasiequilibrium, but also guarantee smallness
of the angle between the direction $f-\Pi^*(f)$ and tangent space to
$\mathfrak{q}$.

One could rely on the stability of this direction, but we fail to
prove this in any general case. The directional instability changes
the structure of dissipation terms: the accuracy decreases to the
first-order in time step $h$ and significant fluctuations of the
Prandtl number and viscosity, etc. may occur. This carries a danger
even without blow-ups; one could conceivably be relying on
nonreliable computational results.

\begin{figure}
\begin{centering}
\includegraphics[width=60mm, height=25mm]{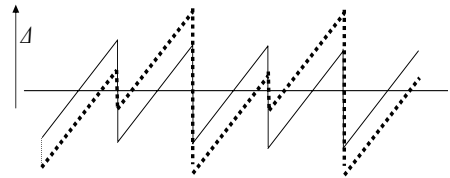}
\caption{\label{NStab}Neutral stability and one-step oscillations in
a sequence of reflections. Bold dotted line -- a perturbed motion,
$\Delta$ -- direction of neutral stability.}
\end{centering}
\end{figure}

\paragraph{Direction of neutral stability.} Further, there exists a neutral stability of all described
approximations that causes one-step oscillations: a small shift of
$f$ in the direction of $\Delta_{f^*_M}$ does not relax back for
$\beta=1$, and its relaxation is slow for $\beta \sim 1$ (for small
viscosity). This effect is demonstrated for a chain of mirror
reflections in Fig.~\ref{NStab}.

\subsection{Dissipative recipes for stabilisation}

\begin{figure}
\begin{centering}
\includegraphics[width=60mm, height=32mm]{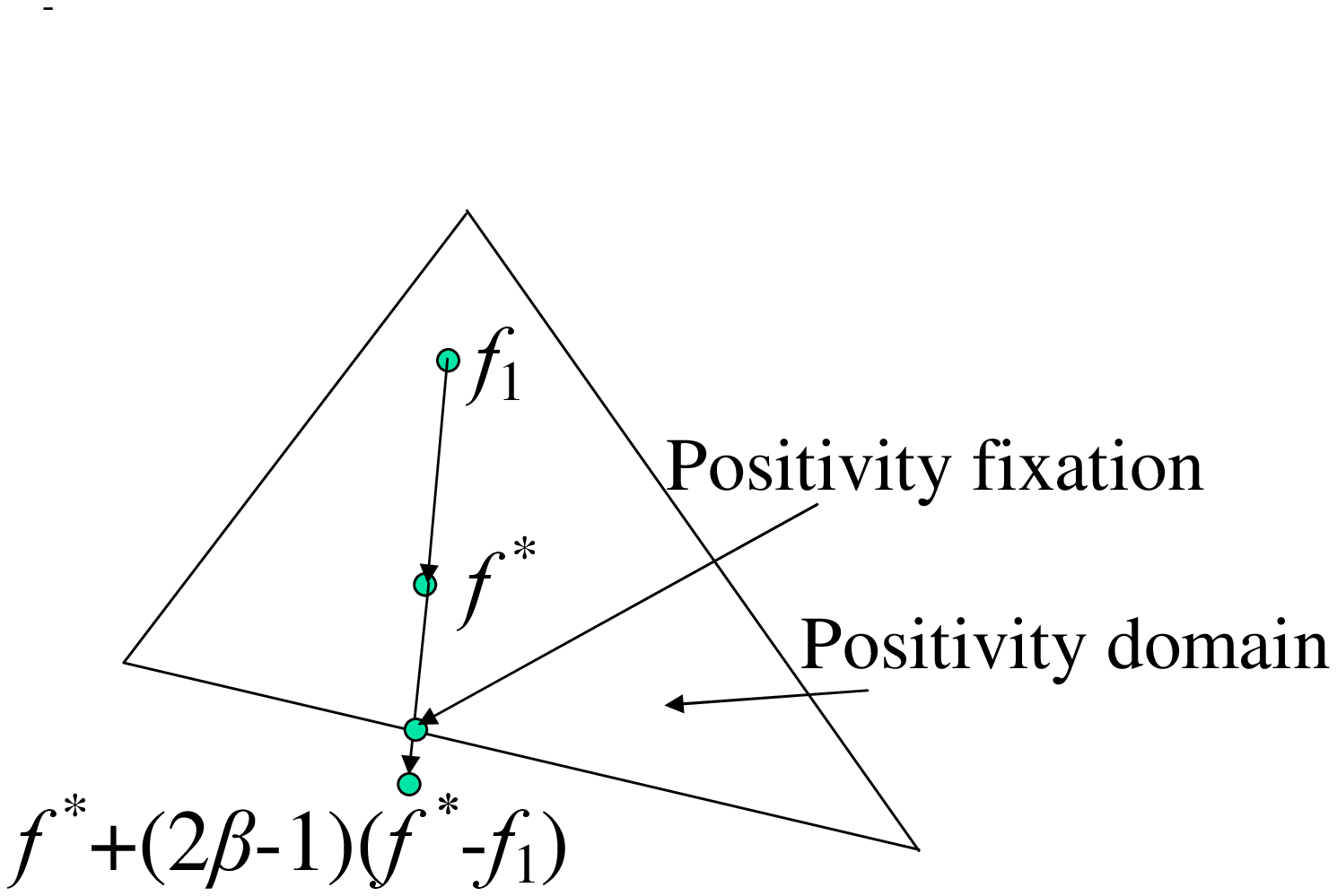}
\caption{\label{PosRule}Positivity rule in action. The motions
stops at the positivity boundary.}
\end{centering}
\end{figure}

\paragraph{Positivity rule.}
There is a simple recipe for positivity preservation
\cite{Rob_preprint,Tosi06}: to substitute nonpositive
$I_0^{\beta}(f)(\xx)$ by the closest nonnegative state that belongs
to the straight line
\begin{equation}\label{StrLine}
  \Bigr\{\lambda f(\xx) + (1-\lambda) \Pi^*(f(\xx))|\: \lambda \in
\mathbb{R}\Bigl\}
\end{equation}
defined by the two points, $f(\xx)$ and its corresponding
quasiequilibrium state. This operation is to be applied point-wise,
at the points of the lattice where the positivity is violated. The
coefficient $\lambda$ depends on $\xx$ too. Let us call this recipe
the \textit{positivity rule} (Fig.~\ref{PosRule}); it preserves
positivity of populations and probabilities, but can affect the
accuracy of approximation. The same rule is necessary for
ELBGK~\eqref{elbm} when a positive ``mirror state" $\tilde{f}$ with
the same entropy as $f$ does not exists on the straight line
\eqref{StrLine}.

The positivity rule saves the existence of positive solutions, but
affects dissipation because the result of the adjusted collision is
closer to quasiequilibrium. There is a family of methods that modify
collisions at some points by additional shift in the direction of
quasiequilibrium. The positivity rule represents the minimal
necessary modification. It is reasonable to always use this rule for
LBM (as a ``salvation rule").

\paragraph{Ehrenfests' regularisation.}

To discuss methods with additional dissipation, the entropic
approach is very convenient. Let entropy $S(f)$ be defined for each
population vector $f=(f_i)$ (below, we use the same letter $S$ for
local-in-space entropy, and hope that the context will make this
notation clear). We assume that the global entropy is a sum of local
entropies for all sites. The local nonequilibrium entropy is
\begin{equation}\label{locaNEQentropy}
 \Delta S(f)= S(f^*) - S(f),
\end{equation}
where $f^*$ is the corresponding local quasiequilibrium at the same
point.

The {\it Ehrenfests' regularisation}~\cite{Rob_preprint,Rob}
provides ``entropy trimming": we monitor local deviation of $f$ from
the corresponding quasiequilibrium, and when $\Delta S(f,\xx)$
exceeds a pre-specified threshold value $\delta$, perform local
Ehrenfests' steps to the corresponding equilibrium. So that the
Ehrenfests' steps are not allowed to degrade the accuracy of LBGK it
is pertinent to select the $k$ sites with highest $\Delta S>\delta$.
The a posteriori estimates of added dissipation could easily be
performed by analysis of entropy production in Ehrenfests' steps.
Numerical experiments show (see, e.g.,~\cite{Rob_preprint,Rob} and
Sect.~\ref{sec5}) that even a small number of such steps drastically
improves stability.

To avoid the change of accuracy order ``on average", the number of
sites with this step should be $\mathcal{O}(N \delta_x /L)$ where
$N$ is the total number of sites, $\delta_x$ is the step of the
space discretization and $L$ is the macroscopic characteristic
length. But this rough estimate of accuracy in average might be
destroyed by concentrations of Ehrenfests' steps in the most
nonequilibrium areas, for example, in boundary layers. In that case,
instead of the total number of sites $N$ in the estimate
$\mathcal{O}(N \delta_x /L)$ we should take the number of sites in a
specific region~\footnote{We are grateful to I.V.~Karlin for this
comment.}. The effects of concentration could be easily analysed a
posteriori.

\paragraph{Entropic steps for nonentropic equilibria.}

If the approximate discrete equilibrium $f^*$ is nonentropic,  we
can use $\Delta S_{\rm K}(f)=-S_{\rm K}(f)$ instead of $\Delta
S(f)$, where $S_{\rm K}$ is the Kullback entropy. This entropy,
\begin{equation}\label{Kulback}
S_{\rm K}(f)=- \sum_i f_i \ln\biggl(\frac{f_i}{f_i^*}\biggr),
\end{equation}
gives the physically reasonable entropic distance from equilibrium,
if the supposed continuum system has the classical BGS entropy. In
thermodynamics, the Kullback entropy belongs to the family of
Massieu--Planck--Kramers functions (canonical or grand canonical
potentials). One can use~\eqref{Kulback} in the construction of
Ehrenfests' regularisation for any choice of discrete equilibrium.

We have introduced two procedures: the positivity rule and
Ehrenfests' regularisation. Both improve stability, reduce
nonequilibrium entropy, and, hence, nonequilibrium fluxes. The
proper context for discussion of such procedures are the
flux-limiters in finite difference and finite volume methods. Here
we refer to the classical flux-corrected transport (FCT)
algorithm~\cite{BorisBook1} that strictly maintains positivity, and
to its further developments~\cite{BookBoris2,Toro,FCTMono}.

\paragraph{Smooth limiters of nonequilibrium entropy.}

The positivity rule and Ehrenfests' regularisation provide rare,
intense and localised corrections. Of course, it is easy and also
computationally cheap to organize more gentle transformations with
smooth shifts of higher nonequilibrium states to equilibrium. The
following regularisation transformation distributes its action
smoothly:
\begin{equation}\label{smoothReg}
f \mapsto f^*+ \phi (\Delta S(f))(f-f^*).
\end{equation}
The choice of the function $\phi$ is highly ambiguous, for example,
$\phi=1/(1+ \alpha \Delta S^k)$ for some $\alpha>0$, $k>0$. There
are two significantly different choices: (i) ensemble-independent
$\phi$ (i.e., the value of $\phi$ depends on the local value of
$\Delta S$ only) and (ii) ensemble-dependent $\phi$, for example
\begin{equation*} \phi=\frac{1+(\Delta S/(\alpha {\rm E} (\Delta
S)))^{k-1/2}} {1+(\Delta S/(\alpha {\rm E} (\Delta S)))^{k}}
\end{equation*}
where ${\rm E} (\Delta S)$ is the average value of $\Delta S$ in the
computational area, $k\geq 1$ and $\alpha \gtrsim 1$.  It is easy to
select an ensemble-dependent $\phi$ with control of total additional
dissipation.

\paragraph{ELBGK collisions as a smooth limiter.}

On the basis on numerical tests, the authors of~\cite{Tosi06} claim
that the positivity rule provides the same results (in the sense of
stability and absence/presence of spurious oscillations) as the
ELBGK models, but ELBGK provides better accuracy.

For the formal definition of ELBGK~\eqref{elbm} our tests do not
support claims that ELBGK erases spurious oscillations (see
Sect.~\ref{sec5} below). Similar observations for Burgers equation
has been reported in~\cite{Bruce}. We understand this situation in
the following way. The entropic method consists of at least three
components:
\begin{enumerate}
\item{entropic quasiequilibrium defined by entropy
maximisation;}
\item{entropy balanced collisions~\eqref{elbm} that
have to provide proper entropy balance;}
\item{a method for the solution of the transcendental equation $S(f)=S(\tilde{f})$ to find
$\alpha=\alpha(f)$ in~\eqref{elbm}.}
\end{enumerate}
It appears that the first two items do not affect spurious
oscillations at all, if we solve equation for $\alpha(f)$ with
high accuracy. Additional viscosity could, potentially, be added
by use of explicit analytic formulas for $\alpha(f)$. In order not
to decrease entropy, the errors in these formulas always increase
dissipation. This can be interpreted as a hidden transformation of
the form~\eqref{smoothReg}, where the coefficients of $\phi$ also
depend on $f^*$.

Compared to flux limiters, nonequilibrium entropy limiters have a
great benefit: by summation of all entropy changes we can estimate
the amount of additional dissipation the limiters introduce into the
system.

\subsection{Non-dissipative recipes for stabilisation}
\label{SSnonDisRec}
\paragraph{Microscopic error and macroscopic accuracy.} The invariant film $\mathfrak{q}$ is an invariant manifold for the
free-flight transformation and for the temporal and entropic
involutions. The linear involution $I_0$, as well as the ELBGK
involution $I_E$, transforms a point $f\in \mathfrak{q}$ into a
point $f'$ with $f-f' = \alpha \Delta_{\Pi^*(f)} + o(f-\Pi^*(f))$,
i.e., the vector $f-f'$ is ``almost tangent" to $\mathfrak{q}$, and
the distance from $f'$ to $\mathfrak{q}$ has the order
$\mathcal{O}(\| f-\Pi^*(f) \|^2)$.

Hence, if the initial state belongs to $\mathfrak{q}$, and the
distance from quasiequilibrium is small enough ($\sim \mathcal{O}
(h)$), then during several steps the LBGK chain will remain near
$\mathfrak{q}$ with deviation $\sim \mathcal{O}(h^2)$. Moreover,
because errors produced by collisions (deviations from
$\mathfrak{q}$) have zero macroscopic projection, the corresponding
macroscopic error in $M$ during several steps will remain of order
$\mathcal{O}(h^3)$.

To demonstrate this, suppose the error in $f$, $\delta f$, is of
order $\mathcal{O}(h^k)$, and $m(\delta f)= 0$, then for smooth
fields after a free-flight step an error of higher order appears in
the macroscopic variables $M$: $m(\Theta_h (\delta f))=\mathcal{O}
(h^{k+1})$, because $m(\Theta_h (\delta f))= m((\Theta_h-1) (\delta
f))$ and $\Theta_h-1 =\mathcal{O} (h)$. The last estimate requires
smoothness.

This simple statement is useful for the error analysis we perform.
We shall call it the {\it lemma of higher macroscopic accuracy}: a
microscopic error of order $\mathcal{O}(h^k)$ induces, after a time
step $h$, a macroscopic error of order $\mathcal{O}(h^{k+1})$, if
the field of macroscopic fluxes is sufficiently small (here, the
microscopic error means the error that has zero macroscopic
projection).

\paragraph{Restarts and approximation of $\Delta_{f^*_M}$.}

The problem of nondissipative LBM stabilisation we interpret as a
problem of appropriate restart from a point that is sufficiently
close to the invariant film. If $h=\tau$ and collisions return the
state to quasiequilibrium, then the state belongs to $\mathfrak{q}$
for all time with high accuracy. For $h \neq \tau$, formulas for
restarting are also available: one can choose between~\eqref{incond}
and, more flexibly,~\eqref{incond1}. Nevertheless, many questions
remain. Firstly, what should one take for $\Delta_{f^*_M}$? This
vector has a straightforward differential definition~\eqref{defect}
(let us also recall that $\tau \Delta_{f^*_M}$ is the first
Chapman--Enskog nonequilibrium correction to the distribution
function~\eqref{firstChEnBGKmic}). But numerical differentiation
could violate the exact-in-space free-flight transformation and
local collisions. There exists a rather accurate central difference
approximation of $\Delta_{f^*_M}$ on the basis of free-flight:
\begin{equation}\label{DeltaCentral}
\Delta_{f^*_M}=\frac{1}{2}(\Delta^+_{f^*_M}+\Delta^-_{f^*_M}) +
\mathcal{O}(h^2),
\end{equation}
where
\begin{align*}
\Delta^+_{f^*_M}&=\frac{1}{h}(\Theta_h(f^*_M)-\Pi^*(\Theta_h(f^*_M))),
\\
\Delta^-_{f^*_M}&=-\frac{1}{h}(\Theta_{-h}(f^*_M)-\Pi^*(\Theta_{-h}(f^*_M))).
\end{align*}
There are no errors of the first-order in~\eqref{DeltaCentral}. The
forward ($\Delta^+_{f^*_M}$) and backward ($\Delta^-_{f^*_M}$)
approximations are one order less accurate. The computation of
$\Delta^\pm_{f^*_M}$ is of the same computational cost as an LBGK
step, hence, if we use the restart formula~\eqref{incond} with
central difference evaluation of
$\Delta_{f^*_M}$~\eqref{DeltaCentral}, then the computational cost
increases three times (approximately). Non-locality of collisions
(restart from the distribution $f^s_M$~\eqref{incond} with a
nonlocal expression for $\Delta_{f^*_M}$) spoils the main LBM idea
of exact linear free-flight and local collisions: nonlocality is
linear and exact, nonlinearity is local~\cite{succi01}). One might
also consider the inclusion of other finite difference
representations for $\Delta_{f^*_M}$ into explicit LBM schemes. The
consequences of this combination should be investigated.

\paragraph{Coupled steps with quasiequilibrium ends.}

The mean viscosity lemma allows us to combine different starting
points in order to obtain the necessary macroscopic equations. From
this lemma, it follows that the following construction of two
coupled steps with restart from quasiequilibrium approximates the
macroscopic equation~\eqref{MACRO1} with second-order accuracy in
time step $h$.

Let us take $f^*_M$ as the initial state with given $M$, then evolve
the state by $\Theta_{h}$, apply the incomplete temporal involution
$I_T^{\beta}$~\eqref{TempInvInc}, again evolve by $\Theta_{h}$, and
finally project by $\Pi^*$ onto the quasiequilibrium manifold:
\begin{equation}\label{Tstabcoupst}
M\mapsto  M'= m(\Pi^*(\Theta_{h}(I_T^{\beta} (\Theta_{h}(f^*_M))))).
\end{equation}
It follows from the restart formula~\eqref{incond} and the mean
viscosity lemma that this step gives a second-order in time $h$
approximation to the shift in time $2h$ for~\eqref{MACRO1} with
$\tau=2(1-\beta)h$, $1/2\leq \beta\leq 1$. Now, let us replace
$I_T^{\beta}$ by the much simpler transformation of LBGK collisions
$I_0^{\beta}$~\eqref{linInv}:
\begin{equation}\label{stabcoupst}
M\mapsto M'= m(\Pi^*(\Theta_{h}(I_0^{\beta} (\Theta_{h}(f^*_M))))).
\end{equation}
\begin{figure}[h]
\begin{centering}
\includegraphics[width=70mm, height=32mm]{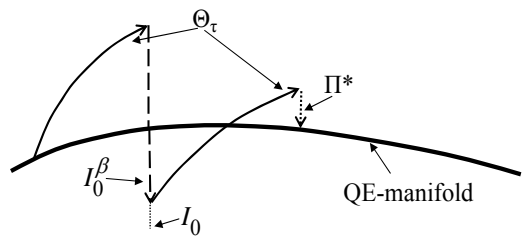}
\caption{\label{CoupledStep}The scheme of coupled steps
\eqref{stabcoupst}.\label{fig2}}
\end{centering}
\end{figure}
According to the lemma of higher macroscopic accuracy this step
(Fig.~\ref{CoupledStep}) also gives a second-order in time $h$
approximation to the shift in time $2h$ for~\eqref{MACRO1} with
$\tau=2(1-\beta)h$, $1/2\leq \beta\leq 1$. The replacement of
$I_T^{\beta}$ by $I_0^{\beta}$ introduces an error in $f$ that is of
order $\mathcal{O}(h^2)$, but both transformations conserve the
value of macroscopic variables exactly. Hence (due to the lemma of
higher macroscopic accuracy) the resulting error of coupled steps
\eqref{stabcoupst} in the macroscopic variables $M$ is of order
$\mathcal{O}(h^3)$. This means that the method has second-order
accuracy.

Let us enumerate the macroscopic states in~\eqref{stabcoupst}:
$M_0=M$, $M_{1/2}=m(\Theta_{h}(f^*_M))$ and $M_1=M'$.  The shift
from $M_0$ to $M_{1/2}$ approximates the shift in time $h$
for~\eqref{MACRO1} with $\tau=h$. If we would like to
model~\eqref{MACRO1} with $\tau \ll h$, then $\tau=h$ means
relatively very high viscosity. The step from $M_{1/2}$ to $M_1$ has
to normalize viscosity to the requested small value (compare to
\textit{antidiffusion} in~\cite{BorisBook1,BookBoris2}). The
antidiffusion problem necessarily appears in most CFD approaches to
simulation flows with high-Reynolds numbers. Another famous example
of such a problem is the filtering-defiltering problem in large eddy
simulation (LES)~\cite{Stolz}. The antidiffusion in the coupled
steps is produced by physical fluxes (by free-flight) and preserves
positivity. The coupled step is a transformation $M_0 \mapsto M_1$
and takes time $2h$. The middle point $M_{1/2}$ is an auxiliary
state only.

Let us enumerate the microscopic states in~\eqref{stabcoupst}:
$f_0=f^*_M$, $f^-_{1/2} =\Theta_{h}(f^*_M)$,
$f^0_{1/2}=\Pi^*(f^-_{1/2})$, $\tilde{f}_{1/2}=I_0^{1}(f^-_{1/2})$,
$f^+_{1/2}  = f^0_{1/2}+ (1- \beta)(\tilde{f}_{1/2}-f^0_{1/2})$,
$f_1^- = \Theta_{h} (f^+_{1/2})$ $f_1 = \Pi^*(f_1^-)= f^*_{M_1}$,
where $M_1=m(f_1^-)$. Here, in the middle of the step, we have 4
points: a free-flight shift of the initial state ($f^-_{1/2}$), the
corresponding quasiequilibrium ($f^0_{1/2}$), the mirror image
($\tilde{f}_{1/2}$) of the point $f^-_{1/2}$ with respect to the
centre $f^0_{1/2}$, and the state ($f^+_{1/2}$) that is the image of
$f^-_{1/2}$ after homothety with centre $f^0_{1/2}$ and coefficient
$2\beta-1$.

For smooth fields, the time shift $\Theta_{h}$ returns
$\tilde{f}_{1/2}$ to the quasiequilibrium manifold with possible
error of order $\mathcal{O}(h^2)$. For entropic equilibria, the
nonequilibrium entropy of the state $\Theta_{h}(\tilde{f}_{1/2})$ is
of order $\mathcal{O}(h^4)$. This is an entropic estimate of the
accuracy of antidiffusion: the nonequilibrium entropy of
$\tilde{f}_{1/2}$ could be estimated from below as $C(M)h^2$, where
$C(M)>0$ does not depend on $h$. The problem of antidiffusion can be
stated as an implicit stepping problem: find a point $\tilde{f}$
such that
\begin{equation}\label{antidiff}
m(\tilde{f})=M,\quad (\Pi^*-1)(\Theta_h(\tilde{f}))=0.
\end{equation}
This {\it antidiffusion problem} is a proper two-point boundary
value problem. In a finite-dimensional space the first condition
includes $N$ independent equations (where $N$ is the number of
independent macroscopic variables), the second allows $N$ degrees of
freedom, because the values of the macroscopic variables at that end
are not fixed and $\Theta_h(\tilde{f})$ could be any point on the
quasiequilibrium manifold). Shooting methods for the solution of
this problem looks quite simple:
\begin{itemize}
\item{Method A,
\begin{equation}\label{shootA}
\tilde{f}_{n+1}
  =\tilde{f}_{n}+\Pi^*(\Theta_{h}(\tilde{f}_{n})) -
  \Theta_{h}(\tilde{f}_{n}),
\end{equation}}
\item{Method B,
\begin{equation}\label{shootB}
\begin{split}
\tilde{f}_{n+1} &=\Theta_{-h}(\Pi^*(\Theta_{h}(\tilde{f}_{n})))
\\&+f^*_M -\Pi^*(\Theta_{-h}(\Pi^*(\Theta_{h}(\tilde{f}_{n})))).
\end{split}
\end{equation}}
\end{itemize}

Method A is: shoot from the previous approximation, $\tilde{f}_{n}$,
by $\Theta_{h}$, project onto quasiequilibrium,
$\Pi^*(\Theta_{h}(\tilde{f}_{n}))$, and then correction of
$\tilde{f}_{n}$ by the final point displacement,
$\Pi^*(\Theta_{h}(\tilde{f}_{n})) - \Theta_{h}(\tilde{f}_{n})$. The
value of $M$ does not change, because
$m(\Pi^*(\Theta_{h}(\tilde{f}_{n}))) =
m(\Theta_{h}(\tilde{f}_{n}))$.

Method B is: shoot from the previous approximation, $\tilde{f}_{n}$,
by $\Theta_{h}$, project onto quasiequilibrium, shoot backwards by
$\Theta_{-h}$, and then correction of $M$ using quasiequilibria
(plus the quasiequilibrium with required value of $M$, and minus one
with current value of $M$).

The initial approximation could be $\tilde{f}_{1/2}$, and $n$ here
is the number of iteration. Due to the lemma of higher macroscopic
accuracy, each iteration~\eqref{shootA} or~\eqref{shootB} increases
the order of accuracy (see also the numerical test in
Sec.~\ref{sec5}).

The shooting method A~\eqref{shootA} better meets the main LBM idea:
each change of macroscopic variable is due to a free-flight step
(because free-flight in LBM is exact), all other operations effect
nonequilibrium component of the distribution only. The correction of
$M$ in the shooting method B~\eqref{shootB} violates this
requirement.

The idea that all macroscopic changes are projections of free-flight
plays, for the proposed LBM antidiffusion, the same role as the
monotonicity condition for FCT~\cite{BorisBook1}. In particular,
free-flight never violates positivity.

If we a find solution $\tilde{f}$ to the antidiffusion problem with
$M=M_{1/2}$, then we can take $f^+_{1/2}  = f^0_{1/2}+ (1-
\beta)(\tilde{f}-f^0_{1/2})$, and $M_1=m(\Theta_h (f^+_{1/2}))$. But
even exact solutions of~\eqref{antidiff} can cause stability
problems: the entropy of $\tilde{f}$ could be less than the entropy
of $f^-_{1/2}$, and blow-up could appear. A palliative solution is
to perform an entropic step: to find $\alpha$ such that
$S(f^0_{1/2}+ \alpha (\tilde{f}-f^0_{1/2}))=S(f^-_{1/2})$, then use
$f^+_{1/2}  = f^0_{1/2}+ (1- \beta)\alpha(\tilde{f}-f^0_{1/2})$.
Even for nonentropic equilibria it is possible to use the Kullback
entropy~\eqref{Kulback} for comparison of distributions with the
same value of the macroscopic variables. Moreover, the quadratic
approximation to~\eqref{Kulback} will not violate second-order
accuracy, and does not require the solution of a transcendental
equation.

The viscosity coefficient is proportional to $\tau$ and
significantly depends on the chain construction: for the
sequence~\eqref{coup} we have $\tau=(1-\beta)h / \beta$, and for the
sequence of steps~\eqref{stabcoupst} $\tau=2(1-\beta)h$. For small
$1-\beta$ the later gives around two times larger viscosity (and for
realisation of the same viscosity we must take this into account).

How can the coupled steps method~\eqref{stabcoupst} fail? The method
collects all the high order errors into dissipation. When the
high-order errors accumulated in dissipation become compatible with
the second-order terms, the observable viscosity significantly
increases. In our numerical tests this catastrophe occurs when the
hydrodynamic fields change significantly on 2-3 grid steps
$\delta_x$ (the characteristic wave length $\lambda \sim
3\delta_x$). The catastrophe point is the same for the plain coupled
steps~\eqref{stabcoupst} and for the methods with iterative
corrections~\eqref{shootA} or~\eqref{shootB}. The appropriate
accuracy requires $\lambda \gtrsim 10\delta_x$. On the other hand,
this method is a good solver for problems with shocks (in comparison
with standard LBGK and ELBGK) and produces shock waves with very
narrow fronts and almost without Gibbs effect. So, for sufficiently
smooth fields it should demonstrate second-order accuracy, and in
the vicinity of steep velocity derivatives it increases viscosity
and produces artificial dissipation. Hence, this recipe is
nondissipative in the main order only.

The main technical task in ``stabilisation for accuracy" is to keep
the system sufficiently close to the invariant film. Roughly
speaking, we should correct the microscopic state $f$ in order to
keep it close to the invariant film or to the tangent straight line
$\{f^*_M+ \lambda \Delta_{f^*_M}|\: \lambda \in \mathbb{R}\}$, where
$M=m(f)$. But the general accuracy condition~\eqref{incond1} gives
much more freedom: the restart point should return to the invariant
film in projections on the macroscopic variables and their fluxes
only. A variant of such regularisation was de facto proposed and
successfully tested in~\cite{Latt}. In the simplest realisation of
such approaches a problem of ``ghost" variables~\cite{succi01}  can
arise: when we change the restart~\eqref{incond} to~\eqref{incond1},
neither moments, nor fluxes change. The difference is a ``ghost"
vector. At the next step, the introduced ghost component could
affect fluxes, and at the following steps the coupling between ghost
variables and macroscopic moments emerges. Additional relaxation
times may be adjusted to suppress these nonhydrodynamic ghost
variables~\cite{Dellar}.

\paragraph{Compromise between nonequilibrium memory and restart rules.}

Formulas~\eqref{incond} and~\eqref{incond1} prescribe a choice of
restart states $f^s_M$. All memory from previous evolution is in the
macroscopic state, $M$, only. There is no microscopic (or,
alternatively, nonequilibrium) memory. Effects of nonequilibrium
memory for LBM are not yet well studied. For LBGK with
overrelaxation, these effects increase when $\beta$ approaches 1
because relaxation time decreases. We can formulate a hypothesis:
observed sub-grid properties of various LBGK realisations and
modifications for high-Reynolds number are due to nonequilibrium
memory effects.

In order to find a compromise between the restart requirements
\eqref{incond},~\eqref{incond1} and nonequilibrium memory existence
we can propose to choose directions in concordance with
\eqref{incond},~\eqref{incond1}, where the nonequilibrium entropy
field~\eqref{locaNEQentropy} does not change in the restart
procedure. If after a free-flight step we have a distribution $f$
and find a corresponding restart state $f^s_{m(f)}$ due to a global
rule, then for each grid point $\xx$ we can restart from a point
$f^*_{m(f)}+ \alpha(\xx)(f^s_{m(f)}- f^*_{m(f)})$, where $\alpha
(\xx) >0$ is a solution of the constant local nonequilibrium entropy
equation $S(f^*_{m(f)}(\xx)+ \alpha(\xx)(f^s_{m(f)}(\xx)-
f^*_{m(f)}(\xx)))=S(f(\xx))$. This family of methods allows a
minimal nonequilibrium memory -- the memory about local entropic
distance from quasiequilibrium.

\section{Numerical experiment}\label{sec5}

\subsection{Velocities and equilibria}

To conclude this paper we report two numerical experiments conducted
to demonstrate the performance of some of the proposed LBM
stabilisation recipes from Sect.~\ref{sec3}.

We choose velocity sets with entropic equilibria and an $H$-theorem
in order to compare all methods in a uniform setting.

In 1D, we use a lattice with spacing and time step $h=1$ and a
discrete velocity set $\{v_1,v_2,v_3\}:=\{0,-1,1\}$ so that the
model consists of static, left- and right-moving populations only.
The subscript $i$ denotes population (not lattice site number) and
$f_1$, $f_2$ and $f_3$ denote the static, left- and right-moving
populations, respectively. The entropy is $S=-H$, with
\begin{equation*}
 H = f_1 \log(f_1/4)+f_2\log(f_2)+f_3 \log(f_3),
\end{equation*}
(see, e.g.,~\cite{karlin99}) and, for this entropy, the local
quasiequilibrium state $f^*$ is available explicitly:
\begin{align*}
  f_1^{*} &= \frac{2 n}{3} \bigl(2-\sqrt{1+3 u^2}\bigr), \\
  f_2^{*} &= \frac{n}{6} \bigl((3u-1)+2\sqrt{1+3 u^2} \bigr),\\
  f_3^{*} &= -\frac{n}{6} \bigl((3u+1)-2\sqrt{1+3 u^2} \bigr),
\end{align*}
where
\begin{equation*}
  n := \sum_i f_i,\quad u:=\frac{1}{n} \sum_i v_i f_i.
\end{equation*}

In 2D, the realisation of LBGK that we use will employ a uniform
$9$-speed square lattice with discrete velocities $\{ v_i | \,
i=0,1,\ldots 8\}$: $v_0=0$, $v_i =
(\cos((i-1)\pi/{2}),\sin((i-1){\pi}/{2}))$  for $i=1,2,3,4$, $v_i =
\sqrt{2} (\cos((i-5)\frac{\pi}{2}+\frac{\pi}{4}),
\sin((i-5)\frac{\pi}{2}+\frac{\pi}{4}))$ for $i=5,6,7,8$. The
numbering $f_0$, $f_1,\dotsc,f_8$ are for the static, east, north,
west, south, northeast, northwest, southwest and southeast-moving
populations, respectively. As usual, the quasiequilibrium state,
$f^*$, can be uniquely determined by maximising an entropy
functional
\begin{equation*}
S(f) = -\sum_i f_i \log{\Bigl(\frac{f_i}{W_i}\Bigr)},
\end{equation*}
subject to the constraints of conservation of mass and momentum
\cite{ansumali03}:
\begin{equation}\label{maxwellian}
  f_i^* = n W_i \prod_{j=1}^2
  \Bigl(2-\sqrt{1+3u_j^2}\Bigr)\Biggl( \frac{2u_j+\sqrt{1+3u_j^2}}{1-u_j}
\Biggr)^{v_{i,j}}.
\end{equation}
Here, the \textit{lattice weights,} $W_i$, are given
lattice-specific constants: $W_0=4/9$, $W_{1,2,3,4}=1/9$ and
$W_{5,6,7,8}=1/36$. The macroscopic variables are given by the
expressions
\begin{equation*}
  n := \sum_i f_i,\quad (u_1,u_2) := \frac{1}{n}\sum_i v_i f_i.
\end{equation*}

As we are advised in Sect.~\ref{sec3}, in all of the experiments,
we implement the positivity rule.

\subsection{Shock tube}

The $1$D shock tube for a compressible isothermal fluid is a
standard benchmark test for hydrodynamic codes.  Our computational
domain will be the interval $[0,1]$ and we discretize this
interval with $801$ uniformly spaced lattice sites. We choose the
initial density ratio as 1:2 so that for $x\leq 400$ we set
$n=1.0$ else we set $n=0.5$.

\paragraph{Basic test: LBGK, ELBGK and Coupled steps.}

We will fix the kinematic viscosity of the fluid at $\nu=10^{-9}$
We should take $\beta=1/(2\nu+1)\approx 1-2\nu$ for LBGK and ELBGK
(with or without the Ehrenfests' regularisation). Whereas, for the
coupled step regularisation, we should take $\beta=1-\nu$.

The governing equations for LBGK are
\begin{equation}\label{lbgkLB}
    f_i(x+v_i,t+1)\! =\!
    f_i^*(x,t)+(2\beta-1)(f_i^*(x,t)-\!f_i(x,t)).
\end{equation}
For ELBGK~\eqref{elbm} the governing equations are:
\begin{equation}\label{elbmLB}
    f_i(x+v_i,t+1) = (1-\beta)f_i^*(x,t)+\beta \tilde{f}_i(x,t),
\end{equation}
with $\tilde{f}=(1-\alpha)f+\alpha f^*$. As previously mentioned,
the parameter, $\alpha$, is chosen to satisfy a constant entropy
condition. This involves finding the nontrivial root of the
equation
\begin{equation}\label{entropy_estimate}
  S((1-\alpha) f+\alpha f^*) = S(f).
\end{equation}
Inaccuracy in the solution of this equation can introduce
artificial viscosity. To solve~\eqref{entropy_estimate}
numerically we employ a robust routine based on bisection. The
root is solved to an accuracy of $10^{-15}$ and we always ensure
that the returned value of $\alpha$ does not lead to a numerical
entropy decrease. We stipulate that if, at some site, no
nontrivial root of~\eqref{entropy_estimate} exists we will employ
the positivity rule instead.

The governing equations for the coupled step regularisation of
LBGK alternates between classic LBGK steps and equilibration:
\begin{equation}\label{coupLB}
\begin{split}
 &f_i(x+v_i ,t+1)\\
 &=\left\{
    \begin{aligned}
        & \!f_i^*(x,t), \qquad \text{$N_{\mathrm{step}}$ odd,}  \\
        & \!f_i^*(x,t)+(2\beta-1)(f_i^*(x,t)-f_i(x,t)),\,\text{$N_{\mathrm{step}}$ even,}
    \end{aligned}
    \right.
\end{split}
\end{equation}
where $N_{\mathrm{step}}$ is the cumulative total number of time
steps taken in the simulation. For coupled steps, only  the result
of a couple of steps has clear physical meaning: this couple
transforms $f_i^*(x,t)$ that appears at the beginning of an odd step
to $f_i^*(x,t)$ that appears at the beginning of the next odd step.

\begin{figure}[ht]
\includegraphics[width=85mm,height=85mm]{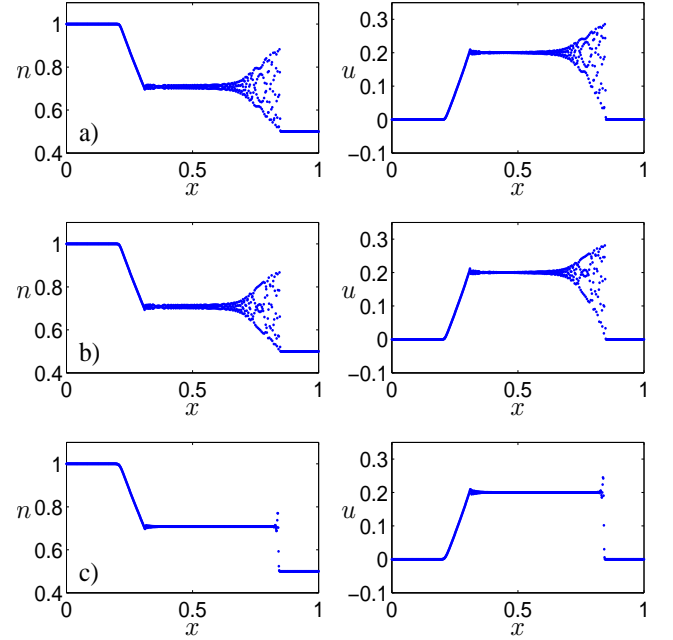}
\caption{Density and velocity profile of the 1:2 isothermal shock
tube simulation after $400$ time steps using (a)
LBGK~\eqref{lbgkLB}; (b) ELBGK~\eqref{elbmLB}; (e) coupled step
regularisation~\eqref{coupLB}; In this example, no negative
population are produced by any of the methods so the positivity rule
is redundant. For ELBGK in this example,~\eqref{entropy_estimate}
always has a nontrivial root. \label{ShokBase}}
\end{figure}

As we can see, the choice between the two collision formulas
LBGK~\eqref{lbgkLB} or ELBGK~\eqref{elbmLB} does not affect spurious
oscillation. But it should be mentioned that the entropic method
consists of not only the collision formula, but, what is important,
includes the provision of special choices of quasiequilibrium that
could improve stability (see, e.g.,~\cite{Shyam2006}). The coupled
steps produce almost no spurious oscillations. This seems to be
nice, but in such cases it is necessary to monitor the amount of
artificial dissipation and to measure the viscosity provided by the
method (see below).

\paragraph{Ehrenfests' regularisation.}

For the realisation of the Ehrenfests' regularisation of LBGK, which
is intended to keep states uniformly close to the quasiequilibrium
manifold, we should monitor nonequilibrium entropy $\Delta S$
\eqref{locaNEQentropy} at every lattice site throughout the
simulation. If a pre-specified threshold value $\delta$ is exceeded,
then an Ehrenfests' step is taken at the corresponding site. Now,
the governing equations become:
\begin{equation}\label{ESLB}
\begin{split}
 &f_i(x+v_i ,t+1)\\
 &=\left\{
    \begin{aligned}
        & \!f_i^*(x,t)+(2\beta-1)(f_i^*(x,t)-f_i(x,t)),\,\text{$\Delta S \leq \delta$,} \\
        & \!f_i^*(x,t), \qquad \text{otherwise,}
    \end{aligned}
    \right.
\end{split}
\end{equation}

Furthermore, so that the Ehrenfests' steps are not allowed to
degrade the accuracy of LBGK it is pertinent to select the $k$ sites
with highest $\Delta S>\delta$. The a posteriori estimates of added
dissipation could easily be performed by analysis of entropy
production in Ehrenfests' steps.

\begin{figure}[ht]
\begin{centering}
\includegraphics[width=85mm,height=56.67mm]{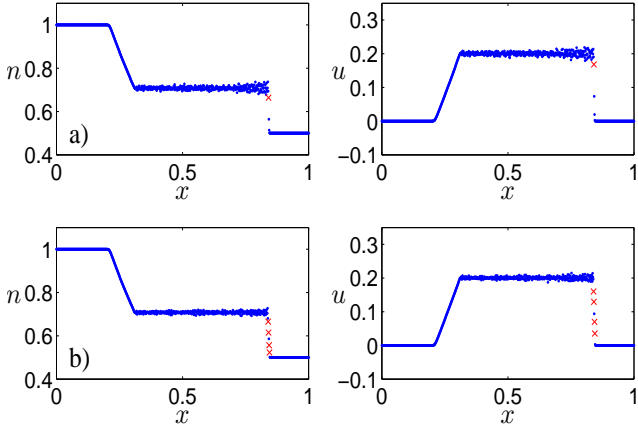}
\caption{\label{shockEhr} Density and velocity profile of the 1:2
isothermal shock tube simulation after $400$ time steps using
Ehrenfests' regularisation~\eqref{ESLB} with (a)
$(k,\delta)=(4,10^{-3})$; (b) $(k,\delta)=(4,10^{-4})$. Sites where
Ehrenfests' steps are employed are indicated by crosses. Compare to
Fig.~\ref{ShokBase}a. \label{EStest1}}
\end{centering}
\end{figure}

\begin{figure}[ht]
\begin{centering}
\includegraphics[width=85mm,height=85mm]{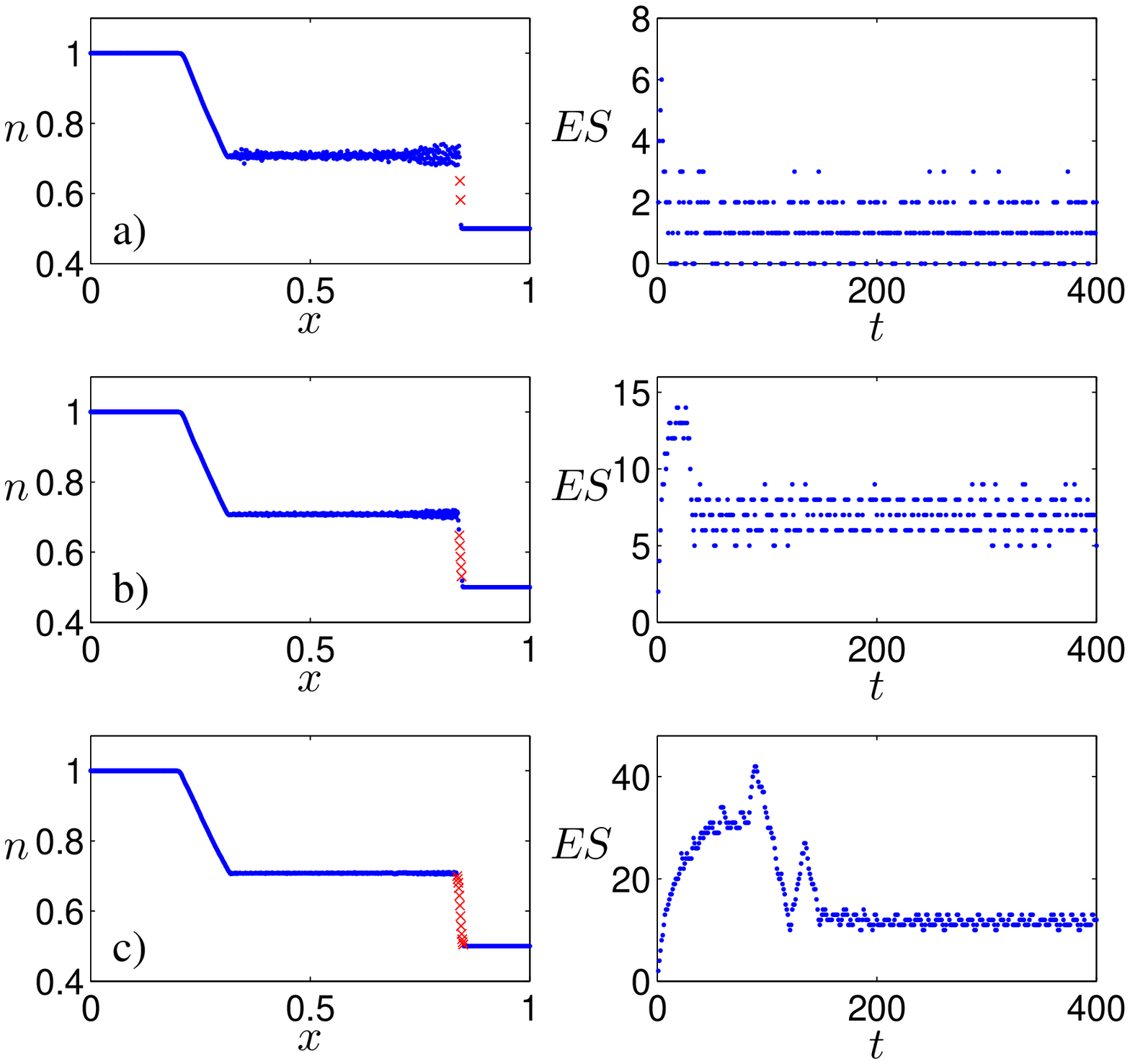}
\caption{LBGK~\eqref{lbgkLB} regularised with Ehrenfests'
steps~\eqref{ESLB}. Density profile of the 1:2 isothermal shock
    tube simulation and Ehrenfests' steps histogram after 400 time steps
    using the tolerances
    (a) $(k,\delta)=(\infty,10^{-3})$;
    (b) $(k,\delta)=(\infty,10^{-4})$; (c) $(k,\delta)=(\infty,10^{-5})$. Sites
    where Ehrenfests' steps are employed are indicated by crosses. Compare to
    Fig.~\ref{ShokBase}a.
    \label{shocktube_unbound_k}\label{fig4}}
\end{centering}
\end{figure}

\begin{figure}[ht]
\begin{centering}
\includegraphics[width=85mm,height=85mm]{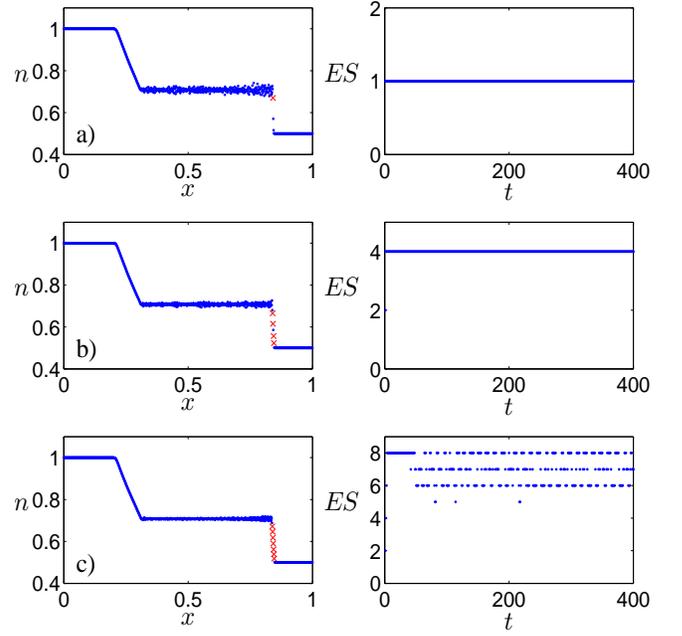}
\caption{LBGK~\eqref{lbgkLB} regularised with Ehrenfests'
steps~\eqref{ESLB}. Density profile of the 1:2 isothermal shock
    tube simulation and Ehrenfests' steps histogram after 400 time steps
    using the tolerances
    (a) $(k,\delta)=(1,10^{-4})$;
    (b) $(k,\delta)=(4,10^{-4})$; (c) $(k,\delta)=(8,10^{-4})$. Sites
    where Ehrenfests' steps are employed are indicated by crosses. Compare to
    Fig.~\ref{ShokBase}a.
    \label{shocktube_bound_k}\label{fig5}}
\end{centering}
\end{figure}

In the example in Fig.~\ref{EStest1}, we have considered fixed
tolerances of $(k,\delta)=(4,10^{-3})$ and $(k,\delta)=(4,10^{-4})$
only. We reiterate that it is important for Ehrenfests' steps to be
employed at only a small share of sites. To illustrate, in
Fig.~\ref{shocktube_unbound_k} we have allowed $k$ to be unbounded
and let $\delta$ vary. As $\delta$ decreases, the number of
Ehrenfests' steps quickly begins to grow (as shown in the
accompanying histograms) and excessive and unnecessary smoothing is
observed at the shock. The second-order accuracy of LBGK is
corrupted. In Fig.~\ref{shocktube_bound_k}, we have kept $\delta$
fixed at $\delta=10^{-4}$ and instead let $k$ vary. We observe that
even small values of $k$ (e.g., $k=1$) dramatically improves the
stability of LBGK.

\subsection{Accuracy of coupled steps}

Coupled steps~\eqref{stabcoupst} give the simplest second--order
accurate stabilization of LBGK. Stabilization is guaranteed by
collection of all errors into dissipative terms. But this monotone
collection of errors could increase the higher order terms in
viscosity. Hence, it seems to be necessary to analyze not only
order of errors, but their values too.

For accuracy analysis of coupled steps we are interested in the
error in the antidiffusion step~\eqref{antidiff}. We analyse one
coupled step for $\beta=1$. The motion starts from a
quasiequilibrium $f_0=f^*_M$, then a free-flight step $f^-_{1/2}
=\Theta_{h}(f^*_M)$, after that a simple reflection
$\tilde{f}_{1/2}=I_0^{1}(f^-_{1/2})$ with respect to the
quasiequilibrium centre $f^0_{1/2}=\Pi^*(f^-_{1/2})$,  again a
free-flight step, $f^-_1=\Theta_{h}(\tilde{f}_{1/2})$, and finally
a projection onto quasiequilibrium, $f_1=\Pi^*(f^-_1)$.

\begin{figure}
\includegraphics[width=85mm,height=85mm]{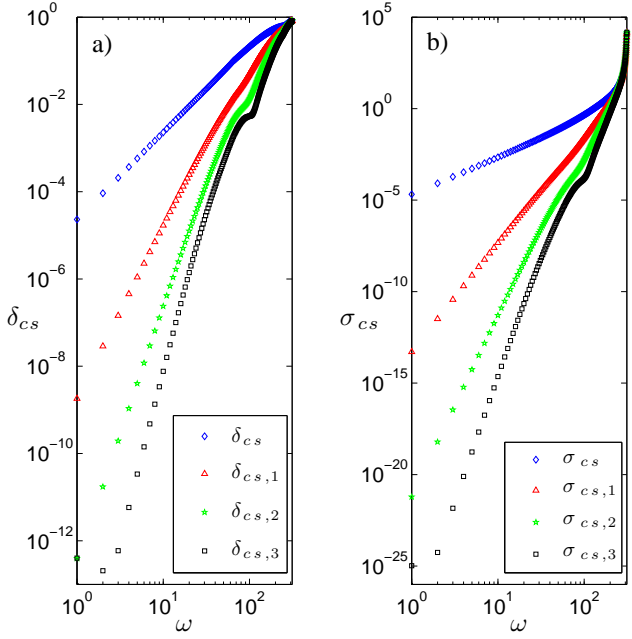}
\caption{\label{AcCoupSt}(a)  The $\ell_2$ estimate of middle point
displacement~\eqref{MidPoinDis}: for coupled steps
\eqref{stabcoupst} (diamonds), one (triangles), two (squares) and
three (dots) shooting iterations~\eqref{shootB}; (b) The $\ell_2$
estimate of nonequilibrity of the final point~\eqref{AddDis}: for
coupled steps~\eqref{stabcoupst} (diamonds), one (triangles), two
(stars) and three (squares) shooting iterations~\eqref{shootB}.}
\end{figure}

In the first of two accuracy tests, two types of errors are to be
studied. The middle point displacement is
\begin{equation}\label{MidPoinDis}
\delta_{\rm cs} = \|f^0_{1/2}  - \Pi^*( \Theta_{-h}(f^-_1))\| /
\|f_0 - f^0_{1/2} \|.
\end{equation}
To estimate nonequilibrity of the final point $f^-_1$ (i.e.,
additional dissipation introduced by last projection in the coupled
step) we should compare the difference $f^-_1 - f_1$ to the
difference at the middle point $f^-_{1/2} - f^0_{1/2}$. Let us
introduce
\begin{equation}\label{AddDis}
\sigma_{\rm cs} = \|f^-_1 - f_1\|^2 / \|f^-_{1/2} - f^0_{1/2} \|^2.
\end{equation}
In our tests (Fig.~\ref{AcCoupSt}) we use the $\ell_2$ norm.

We take the 1D 3-velocity model with entropic equilibria. Our
computational domain will be the interval $[0,1]$ which we
discretize with $1001$ uniformly spaced lattice sites. The initial
condition is $n(x,0)=1+0.2\sin{(2\pi\omega x )}$, $u(x,0)=0.1
\cos{(2\pi\omega x )}$ and we employ periodic boundary conditions.
We compute a single coupled step for frequencies in the range
$\omega=1,2,\dotsc,1000$ (Fig.~\ref{AcCoupSt}).

The solution $\tilde{f}$ to the antidiffusion problem could be
corrected by the shooting iterations~\eqref{shootA} and
\eqref{shootB}. The corresponding errors for method B~\eqref{shootB}
are also presented in Fig.~\ref{AcCoupSt}. We use
$\delta_{\rm{cs},i}$ and $\sigma_{\rm{cs},i}$, for $i=1,2,\dotsc$,
to denote each subsequent shooting of~\eqref{MidPoinDis}
and~\eqref{AddDis}, respectively.

We observe that the nonequilibrity estimate, $\sigma_{\rm cs}$,
blows-up around the wavelength $1/\omega \sim 3\delta_x$.
Simultaneously, the middle point displacement $\delta_{\rm cs}$ has
value around unity at the same point. We do not plot the results for
larger values of $\omega$ as the simulation has become meaningless
and numerical aliasing will now decrease these errors. The same
critical point is observed for each subsequent shooting as well. For
this problem, the shooting procedure is demonstrated to be effective
for wavelengths $1/\omega \lesssim 10\delta_x$.

For the second accuracy test we propose a simple test to measure the
observable viscosity of a coupled step (and LBGK) simulation. We
take the 2D isothermal 9-velocity model with entropic equilibria.
Our computational domain will a square which we discretize with
$(L+1)\times(L+1)$ uniformly spaced points and periodic boundary
conditions. The initial condition is $n(x,y)=1$, $u_1(x,y)=0$ and
$u_2(x,y)=u_0 \sin(2\pi x/L)$, with $u_0=0.05$. The exact velocity
solution to this problem is an exponential decay of the initial
condition: $u_1(x,y,t)=0$, $u_2(x,y,t) = u_0 \exp(-\lambda u_0
t/(\mathrm{Re}L)) \sin(2\pi x/L)$, where $\lambda$ is some constant
and $\mathrm{Re}=\mathrm{Re}(\beta)=u_0 L/\nu(\beta)$ is the
Reynolds number of the flow. Here, $\nu=\nu(\beta)$ is the
theoretical viscosity of the fluid: $\nu=1-\beta$ for the coupled
steps~\eqref{stabcoupst} and $\nu=(1/\beta-1)/2$ for LBGK.

\begin{figure}
\includegraphics[width=85mm,height=56.67mm]{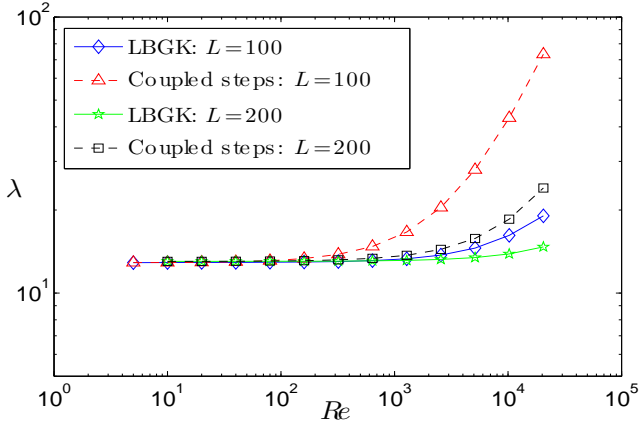}
\caption{Numerically computed value of $\lambda$ versus Reynolds
number for 2D accuracy test: LBGK with $L=100$ (diamonds); coupled
steps~\eqref{stabcoupst} with $L=100$ (triangles); LBGK with $L=200$
(stars); coupled steps~\eqref{stabcoupst} with $L=200$
(squares).\label{AcCoupSt2}}
\end{figure}

Now, we simulate the flow over $L/v_0$ time steps and measure the
constant $\lambda$ from the numerical solution. We do this for both
LBGK and the coupled steps~\eqref{stabcoupst} for $L=100$ and for
$L=200$. The results (Fig.~\ref{AcCoupSt2}) show us that for coupled
steps (and for LBGK to a much lesser extent) the observed viscosity
is higher than the theoretical estimate, hence the observed
$\mathrm{Re}$ is lower than the estimate. In particular, the
lower-resolution ($L=100$) coupled steps simulation diverges from
LBGK at around $\mathrm{Re}=500$. The two times higher-resolution
($L=200$) simulations are close to around
$\mathrm{Re}=\mathcal{O}(1000)$, after which there begins to be a
considerable increase in the observable viscosity (as explained
within Sect.~\ref{SSnonDisRec}).

\subsection{Flow around a square-cylinder}

The unsteady flow around a square-cylinder has been widely
experimentally investigated in the literature (see, e.g.,
\cite{vickery66,davis92,okajima82}). The computational set up for
the flow is as follows. A square-cylinder of side length $L$,
initially at rest, is immersed in a constant flow in a rectangular
channel of length $30L$ and height $25L$. The cylinder is place on
the centre line in the $y$-direction resulting in a blockage ratio
of $4$\%. The centre of the cylinder is placed at a distance
$10.5L$ from the inlet. The free-stream velocity is fixed at
$(u_\infty,v_\infty)=(0.05,0)$ (in lattice units) for all
simulations.

On the north and south channel walls a free-slip boundary condition
is imposed (see, e.g.,~\cite{succi01}). At the inlet, the inward
pointing velocities are replaced with their quasiequilibrium values
corresponding to the free-stream velocity. At the outlet, the inward
pointing velocities are replaced with their associated
quasiequilibrium values corresponding to the velocity and density of
the penultimate row of the lattice.

\paragraph{Maxwell boundary condition.}
The boundary condition on the cylinder that we prefer is the
diffusive Maxwell boundary condition (see,
e.g.,~\cite{cercignani75}), which was first applied to LBM
in~\cite{ansumali02}. The essence of the condition is that
populations reaching a boundary are reflected, proportional to
equilibrium, such that mass-balance (in the bulk) and detail-balance
are achieved. We will describe two possible realisations of the
boundary condition -- time-delayed and instantaneous reflection of
equilibrated populations. In both instances, immediately prior to
the advection of populations, only those populations pointing in to
the fluid at a boundary site are updated. Boundary sites do not
undergo the collisional step that the bulk of the sites are
subjected to.

To illustrate, consider the situation of a wall, aligned with the
lattice, moving with velocity $u_\mathrm{wall}$ and with outward
pointing normal to the wall pointing in the positive $y$-direction
(this is the situation on the north wall of the square-cylinder with
$u_\mathrm{wall}=0$). The time-delayed reflection implementation of
the diffusive Maxwell boundary condition at a boundary site $(x,y)$
on this wall consists of the update
\begin{equation*}
    f_i(x,y,t+1) = \alpha f^{*}_i(u_\mathrm{wall}),\qquad i=2,5,6,
\end{equation*}
with
\begin{equation*}
  \alpha =
  \frac{f_4(x,y,t)+f_7(x,y,t)+f_8(x,y,t)}{f^{*}_2(u_\mathrm{wall})+
  f^{*}_5(u_\mathrm{wall})+f^{*}_6(u_\mathrm{wall})}.
\end{equation*}
Whereas for the instantaneous reflection implementation we should
use for  $\alpha$:
\begin{eqnarray*}
\frac{f_4(x,y+1,t)\!+
f_7(x+1,y+1,t)\!+f_8(x-1,y+1,t)}{f^{*}_2(u_\mathrm{wall})+
  f^{*}_5(u_\mathrm{wall})+f^{*}_6(u_\mathrm{wall})}.
\end{eqnarray*}
Observe that, because density is a linear factor of the
equilibria~\eqref{maxwellian}, the density of the wall is
inconsequential in the boundary condition and can therefore be taken
as unity for convenience.

We point out that, although both realisations agree in the continuum
limit, the time-delayed implementation does not accomplish
mass-balance. Therefore, instantaneous reflection is preferred and
will be the implementation that we employ in the present example.

Finally, it is instructive to illustrate the situation for a
boundary site $(x,y)$ on a corner of the square-cylinder, say the
north-west corner. The (instantaneous reflection) update is then
\begin{equation*}
    f_i(x,y,t+1)= \beta f^{*}_i(u_\mathrm{wall}),\qquad i=2,3,5,6,7,
\end{equation*}
where
\begin{eqnarray*}
\beta &=&\beta_0/\beta_\mathrm{wall}; \\ \beta_0 & =&
f_1(x-1,y,t)+f_4(x,y+1,t)\\ &&+f_5(x-1,y-1,t) +f_7(x+1,y+1,t) \\
&&+f_8(x-1,y+1,t) ;\\
\beta_\mathrm{wall}&=&f^{*}_2(u_\mathrm{wall})+
f^{*}_3(u_\mathrm{wall})\\ &&+f^{*}_5(u_\mathrm{wall})+
  f^{*}_6(u_\mathrm{wall})+f^{*}_7(u_\mathrm{wall}).
\end{eqnarray*}

\paragraph{Strouhal--Reynolds relationship.}
As a test of the Ehrenfests' regularisation~\eqref{ESLB}, a series
of simulations, all with characteristic length fixed at $L=20$, were
conducted over a range of Reynolds numbers $\mathrm{Re}= {L
u_\infty}/{\nu}$. The parameter pair $(k,\delta)$, which control the
Ehrenfests' steps tolerances, are fixed at $(L/2,10^{-3})$.

We are interested in computing the Strouhal--Reynolds relationship.
The Strouhal number $\mathrm{St}$ is a dimensionless measure of the
vortex shedding frequency in the wake of one side of the cylinder:
$\mathrm{St} = {L f_\omega}/{u_\infty},$ where $f_\omega$ is the
shedding frequency.

For our computational set up, the vortex shedding frequency is
computed using the following algorithmic technique. Firstly, the
$x$-component of velocity is recorded during the simulation over
$t_\mathrm{max} = 1250L/u_\infty$ time steps. The monitoring points
is positioned at coordinates $(4L,-2L)$ (assuming the origin is at
the centre of the cylinder). Next, the dominant frequency is
extracted from the final $25$\% of the signal using the discrete
Fourier transform. The monitoring point is purposefully placed
sufficiently downstream and away from the centre line so that only
the influence of one side of the cylinder is recorded.

\begin{figure}
\begin{centering}
\includegraphics[width=85mm,height=56.67mm]{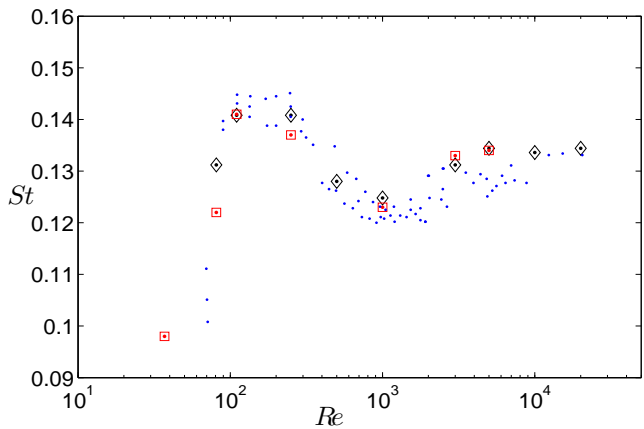}
    \caption{Variation of Strouhal number $\mathrm{St}$ as a function of Reynolds. Dots are Okajima's
    experimental data~\cite{okajima82} (the data has been digitally
extracted from the original paper). Diamonds are the Ehrenfests'
regularisation of LBGK and the squares are the ELBGK simulation
from~\cite{ansumali04}.}\label{fig6}
\end{centering}
\end{figure}

The computed Strouhal--Reynolds relationship using the Ehrenfests'
regularisation of LBGK is shown in Fig.~\ref{fig6}. The simulation
compares well with Okajima's data from wind tunnel and water tank
experiment~\cite{okajima82}. The present simulation extends previous
LBM studies of this problem~\cite{ansumali04,baskar04} which have
been able to quantitively captured the relationship up to
$\mathrm{Re}=\mathcal{O}(1000)$. Fig.~\ref{fig6} also shows the
ELBGK simulation results from~\cite{ansumali04}. Furthermore, the
computational domain was fixed for all the present computations,
with the smallest value of the kinematic viscosity attained being
$\nu = 5\times 10^{-5}$ at $\mathrm{Re}=20000$. It is worth
mentioning that, for this characteristic length, LBGK exhibits
numerical divergence at around $\mathrm{Re}=1000$. We estimate that,
for the present set up, the computational domain would require at
least $\mathcal{O}(10^7)$ lattice sites for the kinematic viscosity
to be large enough for LBGK to converge at $\mathrm{Re}=20000$. This
is compared with $\mathcal{O}(10^5)$ sites for the present
simulation.

\section{Conclusions}\label{sec6}

In this paper, we have analysed LBM as a discrete dynamical system
generated in distribution space by free-flight for time $\delta_t=h$
and involution (temporal or entropic, or just a standard LBGK
reflection that approximates these involutions with second-order
accuracy). Dissipation is produced by superposition of this
involution with a homothety with centre in quasiequilibrium and
coefficient $2\beta -1$.

Trajectories of this discrete dynamical system are projected on to
the space of macroscopic variables, hydrodynamic fields, for
example. The projection of a time step of the LBM dynamics in
distribution space approximates a time shift for a macroscopic
transport equation. We represent the general form of this equation
\eqref{MACRO1}, and provide necessary and sufficient conditions for
this approximation to be of second-order accuracy in the time step
$h$~\eqref{incond},~\eqref{incond1}. This analysis includes
conditions on the free-flight initial state, and does not depend on
the particular collision model.

It is necessary to stress that for free-flight the space
discretization is exact (introduces no errors), if the set of
velocities consists of automorphisms of the grid.

It seems natural to discuss the LBM discrete dynamical system as an
approximate solution to the kinetic equation, for example, to the
BGK kinetics with a discrete velocity set~\eqref{lbgk}. With this
kinetic equation we introduce one more time scale, $\tau$. For $h >
\tau$ (overrelaxation) the discrete LBM does not give a second-order
in time step $h$ approximation to the continuous-in-time equation
\eqref{lbgk}. This is obvious by comparison of ``fast" direction
relaxation times: it is $\tau$ for~\eqref{lbgk} and $h/(2(1-\beta))
\sim h^2/ \tau$ for discrete dynamics (see also~\cite{Xiong}).
Nevertheless, the ``macroscopic shadow" of the discrete LBM with
overrelaxation approximates the macroscopic transport equation with
second-order in time step $h$ accuracy under the conditions
\eqref{incond} and~\eqref{incond1}.

We have presented the main mechanisms of observed LBM
instabilities:
\begin{enumerate}
\item{positivity loss due to high local deviation from
quasiequilibrium;}
\item{appearance of neutral stability in some directions in the zero
viscosity limit;}
\item{directional instability.}
\end{enumerate}
We have found three methods of stability preservation. Two of them,
the positivity rule and the Ehrenfests' regularisation, are
``salvation'' (or ``SOS") operations. They preserve the system from
positivity loss or from the local blow-ups, but introduce artificial
dissipation and it is necessary to control the number of sites where
these steps are applied. In order to preserve the second-order of
LBM accuracy, in average, at least, it is worthwhile to perform
these steps on only a small number of sites; the number of sites
should not be higher than $\mathcal{O}(N\delta_x/L)$, where $N$ is
the total number of sites, $L$ is the macroscopic characteristic
length and $\delta_x$ is the lattice step. Moreover, because these
steps have a tendency to concentrate in the most nonequilibrium
regions (boundary layers, shock layers, etc.), instead of the total
number of sites one can use an estimate of the number of sites in
this region.

The positivity rule and the Ehrenfests' regularisation are members
of a wide family of ``nonequilibrium entropy limiters" that will
play the same role, for LBM, as the flux limiters play for finite
difference, finite volume and finite element methods. We have
described this family and explained how to use entropy estimates for
nonentropic equilibria. The great benefit of the LBM methods is that
the dissipation added by limiters could easily be estimated a
posteriori by summarising the entropy production.

Some practical recommendation for use of nonequilibrium entropy
limiters are as follows:
\begin{itemize}
\item{there exists a huge freedom in the construction of these
limiters;}
\item{for any important class of problems a specific optimal limiter
could be found;}
\item{one of the simplest and computationally cheapest nonequilibrium entropy
limiters is the Ehrenfests' regularisation with equilibration at $k$
sites with highest nonequilibrium entropy $\Delta S > \delta$ (the
$(k,\delta)$-rule);}
\item{the positivity rule should always be implemented.}
\end{itemize}

The developed restart methods (Sec.~\ref{SSnonDisRec}) (including
coupled steps with quasiequilibrium ends) could provide second-order
accuracy, but destroy the memory of LBM. This memory emerges in LBM
with overrelaxation because of slow relaxation of nonequilibrium
degrees of freedom (there is no such memory in the
continuous-in-time kinetic equation with fast relaxation to the
invariant slow Chapman--Enskog manifold). Now, we have no theory of
this memory but can suggest a hypothesis that this memory is
responsible for the LBM sub-grid properties. A compromise between
memory and stability is proposed: one can use the directions of
restart to precondition collisions, and keep the memory in the value
of the field of local nonequilibrium entropy $\Delta S$ (or, for
systems with nonentropic equilibria, in the value of the
corresponding Kullback entropies~\eqref{Kulback}). Formally, this
preconditioning generates a matrix collision model~\cite{succi01}
with a specific choice of matrix: in these models, the collision
matrix is a superposition of projection (preconditioner), involution
and homothety. A return from the simplest LBGK collision to matrix
models has been intensively discussed recently in development of the
multirelaxation time (MRT) models (for example,~\cite{MRT,Du}, see
also~\cite{Rasin} for matrix models for modelling of nonisotropic
advection-diffusion problems, and~\cite{Latt} for regularisation
matrix models for stabilisation at high-Reynolds numbers).

For second-order methods with overrelaxation, adequate second-order
boundary conditions have to be developed. Without such conditions
either additional dissipation or instabilities appear in boundary
layers. The proposed schemes should now be put through the whole
family of tests in order to find their place in the family of the
LBM methods.

Recently, several approaches to stable LBM modelling of
high-Reynolds number flows on coarse grids have been reported
\cite{Geier,Du,KAAOS}. Now it is necessary to understand better the
mechanisms of the LBM sub-grid properties, and to create the theory
that allows us to prove the accuracy of LBM for under-resolved
turbulence modelling.

\acknowledgments

The preliminary version of this work~\cite{Preprint} was widely
discussed, and the authors are grateful to all the scientists who
participated in this discussion. Special acknowledgments goes to
V.~Babu for comments relating to square-cylinder experiments,
H.~Chen who forced us to check numerically the viscosity of the
coupled steps model, S.~Chikatamarla for providing the digitally
extracted data used in Fig.~\ref{fig6}, I.V.~Karlin for deep
discussion of results and background, S.~Succi for very inspiring
discussion and support, M.~Geier, I.~Ginzburg, G.~Hazi,
D.~Kehrwald, M.~Krafczyk, A.~Louis, B.~Shi, and A.~Xiong who
presented to us their related published, and sometimes
unpublished, papers.

This work is supported by Engineering and Physical Sciences Research
Council (EPSRC) grant number GR/S95572/01.

\bibliographystyle{apsrev}

\begin{thebibliography}{10}

\bibitem{ansumali04}
S.~Ansumali, S.~S.~Chikatamarla, C.~E.~Frouzakis, and K.~Boulouchos,
\newblock Entropic lattice {B}oltzmann simulation of the flow past
  square-cylinder,
\newblock {Int. J. Mod. Phys. C} {\bf 15}, 435--445 (2004).

\bibitem{ansumali02}
S.~Ansumali and I.~V.~Karlin,
\newblock Kinetic boundary conditions in the lattice {B}oltzmann method,
\newblock { Phys. Rev. E} {\bf 66}, 026311 (2002).

\bibitem{ansumali03}
S.~Ansumali, I.~V.~Karlin, and H.~C.~Ottinger,
\newblock Minimal entropic kinetic models for hydrodynamics,
\newblock { Europhys. Let.} {\bf 63},  798--804 (2003).

\bibitem{baskar04}
G.~Baskar and V.~Babu,
\newblock Simulation of the unsteady flow around rectangular cylinders using
  the {ISLB} method,
\newblock In: {\em 34th AIAA Fluid Dynamics Conference and Exhibit},
  AIAA--2004--2651 (2004).

\bibitem{Benzi}
R.~Benzi, S.~Succi, and M.~Vergassola,
\newblock The lattice {B}oltzmann-equation -- theory and applications,
\newblock { Phys. Rep.} {\bf 222}, 145--197 (1992).

\bibitem{bgk54}
P.~L.~Bhatnagar, E.~P.~Gross, and M.~Krook,
\newblock A model for collision processes in gases. {I}. {S}mall amplitude
  processes in charged and neutral one-component systems,
\newblock { Phys. Rev.} {\bf 94}, 511--525 (1954).

\bibitem{Bruce}
B.~M.~Boghosian, P.~J.~Love, and J.~Yepez,
\newblock Entropic lattice {B}oltzmann model for {B}urgers equation,
\newblock { Phil. Trans. Roy. Soc. A} {\bf 362}, 1691--1702 (2004).

\bibitem{boghosian01}
B.~M.~Boghosian, J.~Yepez, P.~V.~Coveney, and A.~J.~Wager,
\newblock Entropic lattice {B}oltzmann methods,
\newblock { R. Soc. Lond. Proc. Ser. A }
{\bf 457} (2007), 717--766 (2001).

\bibitem{BookBoris2}
D.~L.~Book, J.~P.~Boris,  and K.~ Hain, Flux corrected transport II.
Generalizations of the method, J. Comput. Phys. {\bf 18}, 248-283
(1975).

\bibitem{BorisBook1}
J.~P.~Boris  and D.~L.~Book, Flux corrected transport I. SHASTA, a
fluid transport algorithm that works, J. Comput. Phys. {\bf 11},
38--69 (1973).

\bibitem{Rob_preprint}
R.~A.~Brownlee, A.~N.~Gorban, and J.~Levesley,
\newblock Stabilisation of the lattice-{B}oltzmann method using the
  {E}hrenfests' coarse-graining,
\newblock { cond-mat/0605359} (2006).

\bibitem{Rob}
R.~A.~Brownlee, A.~N.~Gorban, and J.~Levesley,
\newblock Stabilization of the lattice-{B}oltzmann method using the
  {E}hrenfests' coarse-graining,
\newblock { Phys. Rev. E} {\bf 74}, 037703 (2006).

\bibitem{Preprint}R.~A.~Brownlee, A.~N.~Gorban, and J.~Levesley, Stability and
stabilization of the lattice Boltzmann method: Magic steps and
salvation operations, {cond-mat/0611444} (2006).

\bibitem{Chapman}
S.~Chapman and T.~Cowling,
\newblock {\em Mathematical theory of non-uniform gases}
\newblock  (Third edition, Cambridge University Press, Cambridge, 1970).

\bibitem{cercignani75}
C.~Cercignani,
\newblock {\em Theory and Application of the Boltzmann Equation}
\newblock (Scottish Academic Press, Edinburgh, 1975).

\bibitem{LB1}
S.~Chen and G.~D.~Doolen,
\newblock Lattice Boltzmann method for fluid flows,
\newblock { Annu. Rev. Fluid. Mech.} {\bf 30}, 329--364 (1998).

\bibitem{Shyam2006}S.~S.~Chikatamarla and I.~V.~Karlin,
Entropy and Galilean Invariance of Lattice Boltzmann Theories, {
Phys. Rev. Lett.} {\bf 97}, 190601 (2006)

\bibitem{Raz}A.~J.~Chorin, O.~H.~Hald, and  R.~Kupferman,  Optimal
prediction with memory,  {Physica D} {\bf 166}, 239--257 (2002).

\bibitem{davis92}
R.~W.~Davis and E.~F.~Moore,
\newblock A numerical study of vortex shedding from rectangles,
\newblock { J. Fluid Mech.} {\bf 116}, 475--506 (1982).

\bibitem{Dellar}P.~J.~Dellar, Incompressible limits of lattice Boltzmann equations
using multiple relaxation times, J. of Comput. Phys. {\bf 190},
351–-370 (2003).

\bibitem{Du}R.~Du, B.~Shi, and X.~Chen, Multi-relaxation-time lattice Boltzmann
model for incompressible flow, Phys. Lett. A {\bf 359}, 564-–572
(2006).

\bibitem{ehrenfest11}
P.~Ehrenfest and T.~Ehrenfest,
\newblock {\em The conceptual foundations of the statistical approach in
  mechanics}
\newblock (Dover Publications Inc., New York, 1990).

\bibitem{Geier}
M.~Geier, A.~Greiner, and J.~G.~Korvink, Cascaded digital lattice
Boltzmann automata for high Reynolds number flow, Phys. Rev. E {\bf
73}, 066705 (2006).

\bibitem{gorban06}
A.~N.~Gorban,
\newblock Basic types of coarse-graining,
\newblock In: A.~N.~Gorban, N.~Kazantzis, I.~G.~Kevrekidis, H.~C.~\"{O}ttinger,
  and C.~Theodoropoulos, editors, {\em Model Reduction and Coarse-Graining
  Approaches for Multiscale Phenomena}, 117--176 (Springer,
  Berlin-Heidelberg-New York, 2006)
\newblock cond-mat/0602024.

\bibitem{ocherki}A.~N.~Gorban, V.~I.~Bykov, and G.~S.~Yablonskii,
{\em Essays on Chemical relaxation} (Novosibirsk, Nauka Publ.,
1986).

\bibitem{GKMod}A.~N.~Gorban and I.~V.~Karlin,  General approach to
constructing models of the Boltzmann equation, { Physica A} {\bf
206}, 401--420 (1994).

\bibitem{Plenka}
A.~N.~Gorban and I.~V.~Karlin,
\newblock {\em Geometry of irreversibility: The film of nonequilibrium
states},
\newblock Preprint IHES/P/03/57 (Institut des Hautes \'Etudes Scientifiques,
  Bures-sur-Yvette, France, 2003), \newblock cond-mat/0308331.

\bibitem{GorKar}
A.~N.~Gorban and I.~V.~Karlin,
\newblock {\em Invariant manifolds for physical and chemical kinetics},
vol.  660 of Lect. Notes Phys
\newblock (Springer, Berlin-Heidelberg-New York, 2005).

\bibitem{GorKar2006}A.~N.~Gorban, I.~V.~Karlin, Quasi-Equilibrium Closure Hierarchies
for the Boltzmann Equation, Physica A {\bf 360},  325-–364 (2006).

\bibitem{GKOeTPRE2001}
A.~N.~Gorban, I.~V.~Karlin, H.~C.~\"{O}ttinger, and
L.~L.~Tatarinova,
\newblock Ehrenfest's argument extended to a formalism of nonequilibrium
  thermodynamics,
\newblock { Phys. Rev. E} {\bf 62}, 066124 (2001).

\bibitem{Grad}H.~Grad, On the kinetic theory of rarefied gases, {Comm. Pure and Appl.
Math.} {\bf 2}, 331--407 (1949).

\bibitem{HeLuo}X.~He and L.-S.~Luo, Theory of the lattice Boltzmann method: From
the Boltzmann equation to the lattice Boltzmann equation, Phys. Rev.
E {\bf 56}, 6811--6817 (1997).

\bibitem{Higuera}
F.~Higuera, S.~Succi, and R.~Benzi,
\newblock Lattice gas -- dynamics with enhanced collisions.
\newblock { Europhys. Lett.} {\bf 9}, 345--349 (1989).

\bibitem{Hirt}
C.~W.~Hirt,
\newblock Heuristic stability theory for finite--difference
equations,
\newblock { J. Comput. Phys.} {\bf 2}, 339--355 (1968).

\bibitem{MRT}D.~d~Humi\`eres, I.~Ginzburg, M.~Krafczyk, P.~Lallemand, and
L.-S.~Luo, Multiplerelaxation -- time lattice Boltzmann models in
three-dimensions, Proc. Roy. Soc. London A {\bf 360}, 437--451
(2002).

\bibitem{EIT}D.~Jou, J.~Casas-V\'azquez, G.~Lebon, {\em Extended
irreversible thermodynamics} {(Springer, Berlin, 1993)}.

\bibitem{Junk}M.~Junk, A.~Klar, and L.~S.~Luo, Asymptotic analysis of the lattice
Boltzmann equation, { J. Comput. Phys.} {\bf 210},  676–-704 (2005).

\bibitem{KAAOS}I.~V.~Karlin, S.~Ansumali, E.~De~Angelis,
H.~C.~\"Ottinger, and S.~Succi, Entropic Lattice Boltzmann Method for
Large Scale Turbulence Simulation, { cond-mat/0306003} (2003).

\bibitem{karlin06}
I.~V.~Karlin, S.~Ansumali, C.~E.~Frouzakis, and S.~S.~Chikatamarla,
\newblock Elements of the lattice Boltzmann method I: Linear advection
equation,
\newblock { Commun. Comput. Phys.}, {\bf 1}, 616--655 (2006).

\bibitem{karlin07}I.~V.~Karlin, S.~S.~Chikatamarla, and S.~Ansumali,
\newblock Elements of the lattice Boltzmann method II: Kinetics and hydrodynamics in one
dimension,
\newblock { Commun. Comput. Phys.} {\bf 2}, 196--238 (2007).


\bibitem{karlin99}
I.~V.~Karlin, A.~Ferrante, and H.~C.~\"{O}ttinger,
\newblock Perfect entropy functions of the lattice {B}oltzmann
method,
\newblock { Europhys. Lett.} {\bf 47}, 182--188 (1999).

\bibitem{KGSBPRL}
I.~V.~Karlin, A.~N.~Gorban, S.~Succi, and V.~Boffi,
\newblock Maximum entropy principle for lattice kinetic equations,
\newblock { Phys. Rev. Lett.} {\bf 81}, 6--9 (1998).

\bibitem{FCTMono}D.~Kuzmin, R.~Lohner, and S.~Turek (Eds.), {\em Flux-Corrected
Transport. Principles, Algorithms, and Applications}, (Springer,
Berlin--Heidelberg 2005).

\bibitem{Latt}J.~Latt and B.~Chopard, Lattice Boltzmann method with
regularized pre-collision distribution function, Math. Comput.
Simulat. {\bf 72}, 165--168  (2006).

\bibitem{Lya}A.~M.~Lyapunov, The general problem of the stability of motion (Taylor \& Francis,
London, 1992).

\bibitem{okajima82}
A.~Okajima,
\newblock Strouhal numbers of rectangular cylinders,
\newblock { J. Fluid Mech.} {\bf 123}, 379--398 (1982).

\bibitem{Rasin}I.~Rasin, S.~Succi and W.~Miller, A multi-relaxation lattice kinetic
method for passive scalar diffusion, J. Comput. Phys. {\bf 206},
453--462 (2005).

\bibitem{Robertson}B.~Robertson, Equations of motion in nonequilibrium statistical mechanics, Phys. Rev.,
{\bf 144}, 151--161  (1966).

\bibitem{HudongGrad}X.~Shan, X-F.~Yuan, and H.~Chen, Kinetic theory representation of
hydrodynamics: a way beyond the Navier–Stokes equation, J. Fluid
Mech. {\bf 550}, 413-–441 (2006).

\bibitem{Shokin}
Y.~I.~Shokin
\newblock {\em The method of differential approximation}
\newblock (Springer, New York, 1983).

\bibitem{SChen}J.~D.~Sterling and S.~Chen. Stability analysis of lattice
Boltzmann methods, { J. Comput. Phys.} {\bf 123}, 196--206  (1996).


\bibitem{Stolz}S.~Stolz and N.~A.~Adams, An approximate deconvolution procedure
for large-eddy simulation, Phys. Fluids {\bf 11}, 1699 (1999).

\bibitem{succi01}
S.~Succi,
\newblock {\em The lattice {B}oltzmann equation for fluid dynamics and
beyond}
\newblock (Oxford University Press, New York, 2001).

\bibitem{LB3}
S.~Succi, I.~V.~Karlin, and H.~Chen,
\newblock Role of the {H} theorem in lattice {B}oltzmann hydrodynamic
simulations.
\newblock { Rev. Mod. Phys.} {\bf 74}, 1203--1220 (2002).

\bibitem{Toro}
E.~F.~Toro, {\em Riemann solvers and numerical methods for fluid
dynamics -- a practical introduction} (Springer, Berlin, 1997).

\bibitem{Tosi06}
F.~Tosi, S.~Ubertini, S.~Succi, H.~Chen, and I.~V.~Karlin,
\newblock Numerical stability of entropic versus positivity-enforcing lattice
  {B}oltzmann schemes,
\newblock { Math. Comput. Simulat.} {\bf 72}, 227--231 (2006).

\bibitem{vickery66}
B.~J.~Vickery,
\newblock Fluctuating lift and drag on a long cylinder of square cross-section
  in a smooth and in a turbulent stream,
\newblock { J. Fluid Mech.} {\bf 25}, 481--494 (1966).

\bibitem{Xiong}
A.~Xiong,
\newblock Intrinsic instability of the lattice BGK model,
\newblock { Acta
Mechanica Sinica (English Series)} {\bf 18}, 603--607 (2002).

\end{thebibliography}

\end{document}